\documentclass[12pt]{iopart}
\expandafter\let\csname equation*\endcsname\relax
\expandafter\let\csname endequation*\endcsname\relax
\usepackage[cmex10]{amsmath}
\usepackage[font=footnotesize]{subfig}
\usepackage{amssymb}
\usepackage[dvips]{graphicx}
\usepackage{fixltx2e}
\usepackage[abbr]{harvard}
\usepackage[export]{adjustbox}
\usepackage{textcomp}

\DeclareMathOperator{\SD}{SD}
\DeclareMathOperator{\MAE}{MAE}

\begin{document}
\let\WriteBookmarks\relax
\def\floatpagepagefraction{1}
\def\textpagefraction{.001}

\title{Deep correction of breathing-related artifacts in real-time MR-thermometry}
\author{Baudouin DENIS de SENNEVILLE$^{1,2,3}$, Pierrick COUP\'E$^{4}$, Mario RIES$^{5}$, Laurent FACQ$^{1}$, Chrit MOONEN$^{5}$}
\address{$^1$ University of Bordeaux, IMB, UMR CNRS 5251, Talence, France, F-33405 Talence Cedex, France}
\address{$^2$ INRIA Project team Monc, Talence, France, F-33405 Talence Cedex, France}
\address{$^3$ Department of Radiotherapy, UMC Utrecht, Heidelberglaan 100, 3508 GA, The Netherlands}
\address{$^4$ CNRS, University of Bordeaux, Bordeaux INP, ``Laboratoire Bordelais de la Recherche Informatique'' (LaBRI), UMR5800, F-33400 Talence, France}
\address{$^5$ Imaging Division, UMC Utrecht, Heidelberglaan 100, 3508 GA, Utrecht, The Netherlands}



 
 



\begin{abstract}
Real-time MR-imaging has been clinically adapted for monitoring thermal therapies since it can provide on-the-fly temperature maps simultaneously with anatomical information. However, proton resonance frequency based thermometry of moving targets remains challenging since temperature artifacts are induced by the respiratory as well as physiological motion. If left uncorrected, these artifacts lead to severe errors in temperature estimates and impair therapy guidance.

In this study, we evaluated deep learning for on-line correction of motion related errors in abdominal MR-thermometry. For this, a convolutional neural network (CNN) was designed to learn the apparent temperature perturbation from images acquired during a preparative learning stage prior to hyperthermia. The input of the designed CNN is the most recent magnitude image and no surrogate of motion is needed. During the subsequent hyperthermia procedure, the recent magnitude image is used as an input for the CNN-model in order to generate an on-line correction for the current  temperature map.

The method's artifact suppression performance was evaluated on 12 free breathing volunteers and was found robust and artifact-free in all examined cases. Furthermore, thermometric precision and accuracy was assessed for \textit{in vivo} ablation using high intensity focused ultrasound. All calculations involved at the different stages of the proposed workflow were designed to be compatible with the clinical time  constraints of a therapeutic procedure.
\end{abstract}

\vspace{2pc}

\maketitle

\section{Introduction}

MRI is used for monitoring thermal therapies since it can provide on-line anatomical informations (given by the spatial distribution of the magnitude of the MR-signal) together with temperature mapping \cite{MRgHIFU} \cite{thermo2} \cite{FastThermo}. Many approaches have been developed for MR-thermometry and the Proton-Resonance-Frequency shift (PRF) technique is widely used \cite{magneticFieldChange} \cite{temp_review2} \cite{temp_review}. In the PRF approach, the phase component $\varphi$ of the MR-signal, which is acquired using gradient echo sequences, is directly used to estimate voxel-wise temperature variations \cite{magneticFieldChange} \cite{prf} \cite{prf2}. Due to the spatial phase variations, this signal component needs to be measured on a voxel-per-voxel basis. Let $\vec{r} = (x, y, z) \in \Omega$ be the voxel coordinates, $\Omega$ being the image coordinates domain. An estimate of the temperature change (noted $\Delta T$) at a spatial location $\vec{r}$ and at instant $t$ is obtained by comparing a baseline phase signal acquired at a reference instant $t_{0}$ to the phase signal acquired and at $t$, as follows:

\begin{equation}
\Delta T(\vec{r},t_n) = \left( \varphi (\vec{r},t_0) - \varphi (\vec{r},t) \right) \times k
 \label{eq:PRF}
\end{equation}

\noindent $k$ is a constant parameter, more details on its determination can be found in \cite{temp_coef}. Note that phase wraps need to be compensated on a voxel-by-voxel basis by adding (resp. substracing) $2\pi$ when $\varphi (\vec{r},t_0) - \varphi (\vec{r},t)<-\pi$ (resp. $\varphi (\vec{r},t_0) - \varphi (\vec{r},t)>\pi$).

While this approach works well on static objects, the application of PRF thermometry to moving targets remains challenging since additional variations of the phase component are induced by: (i) moving the observed tissue through an inhomogeneous magnetic field ; (ii) deforming/changing the tissue so that the demagnitisation field of the tissue changes, which are both a consequence of the patient's physiological activity and the associated organ motion \cite{magneticFieldChange}. If left uncorrected, these additional phase variations enter Eq. (\ref{eq:PRF}) in full and could lead to severe thermometric errors, leading to abolute errors exceeding the true temperature difference by more than a magnitude.

As a mitigation strategy, one of the first proposed approaches has been respiratory gating. Respiratory gating consists of intermittent acquisitions performed in each exhalation phase of the respiratory cycle \cite{gating}. As a trigger for the gating, several types of respiratory motion descriptors have been proposed \cite{Resp_model}, ranging from external pressure sensors \cite{gating}, dedicated 1D MR navigator echoes \cite{latency2} to self gated sequences based on MR magnitude images \cite{trackingImf}. Although gating is generally a robust solution to avoid motion induced thermometric errors, it is nevertheless hampered by two drawbacks. First and foremost, the observed motion pattern must be strictly repetitive/periodical and secondly considering a good spatial coverage of the heated region, the achiveable temporal resolution is generally limited to a range of 3 to 6 s \cite{gating1} \cite{gating5} \cite{gating6}. 

In particular the latter motivated the development of non-gated MR-thermometry correction strategies, which are able to selectively remove motion-induced phase changes from the MR-phase and thus to provide artefact-free temperature maps in real-time. However, the required precise modeling of the inhomogeneous magnetic field \textit{in vivo} and the motion associated phase variations, in particular under real-time conditions for therapy guidance, has been difficult to achieve. Most of these early correction strategies can be coarsely classified into two different types, which are generally referred to as ``Referenceless'' and ``Multi-baseline'' PRF thermometry. The interested reader is referred to \cite{PRFCorrComp} for a pragmatic analysis of inherent advantages and drawbacks associated with these two correction strategies:

In referenceless PRF thermometry, the baseline phase signal used to compute the current temperature map is directly estimated from the current MR phase image. To this end, the phase signal of non-heated surrounding tissues is used to extrapolate a baseline phase signal in the targeted area \cite{referenceless} \cite{MRgHIFU2} \cite{referenceless_Rares}. This approach relies on an a priori choice of a region of interest (ROI) and the quality of the thermometry highly depends on an optimal ROI placement. In practice, the fitting ROI has to: (i) encompass --- at least to some extent --- the ablation area ; (ii) be sufficiently close the target area to allow a precise estimate of the background phase there ; (iii) be sufficiently far from the heating zone to be unaffected by heat diffusion and conduction ; (iv) not encompass areas prone to strong local susceptibility variations.
 
\begin{figure}[h!]
\begin{minipage}[b]{\linewidth}
\centering
\centerline{\includegraphics[width=\linewidth]{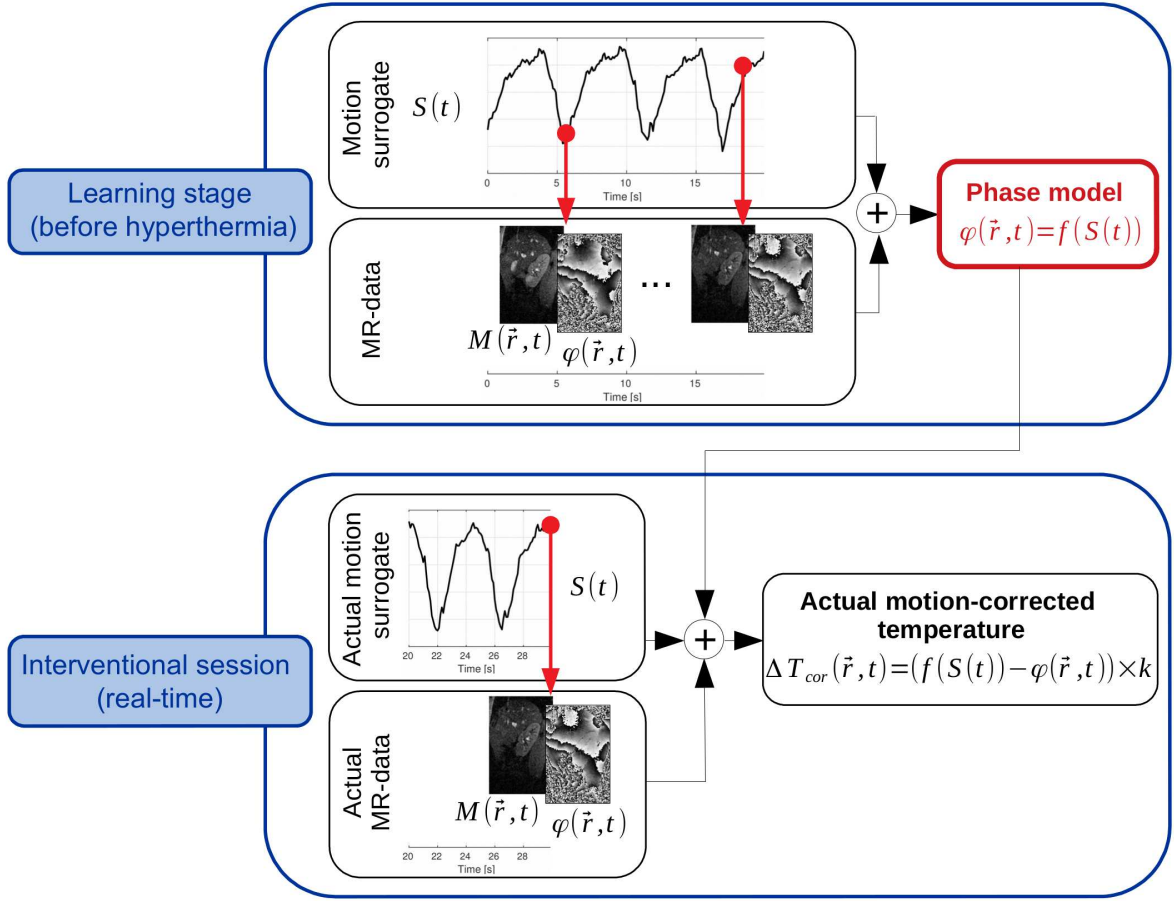}}
\end{minipage}
\caption{Illustration of a typical multi-baseline correction scheme. Both motion surrogate(s) and MR-images are used for this strategy. Both the motion surrogate $S(t)$ and the MR-images are acquired simultaneously in a training phase before the hyperthermia procedure. A multi-baseline collection is used to store each pair of motion surrogate/phase image. During hyperthermia, thermometry artifacts due to the periodical motion of the respiration cycle are addressed by calculating a baseline phase (noted $f(S(t))$) based on the current motion surrogate and the training phase images.}
\label{fig:MB_scheme}
\end{figure}

In multi-baseline PRF thermometry (illustrated in figure \ref{fig:MB_scheme}) a look-up table obtained in absence of temperature variations establishes a relation between the phase variations associated with the patient's physiological motion and a descriptor/detector \cite{gating2}. The benefit of this approach is evident for application scenarios that do not permit a placement of the fitting ROI fulfilling all above-mentioned four conditions simultaneously. This is generally the case for  minimally invasive ablations, or interventions at the boundary of organs. Both MR images and descriptors of motion patterns are continuously and simultaneously acquired during a period covering several respiratory cycles. A look-up table can then be used to store each pair of MR phase image/motion surrogate. During heating, phase artifacts due to the periodical motion of the respiration cycle are addressed by calculating a baseline phase image based on a model of the phase dependence of the current motion descriptor (red block in figure \ref{fig:MB_scheme}). Using multi-baseline strategies, the stability of MR-thermometry largely depends on: (i) the determination of an accurate and precise motion surrogate ; (ii) the accuracy of the model used to address susceptibility related phase changes, especially in regions with complex susceptibility distributions or signal discontinuities. 

More recent approaches proposed to fuse these two largely complementary approaches to combined correction strategies, which compensate the respective weaknesses in order to achieve both increased accuracy and a less convoluted work-flow for clinical applications \cite{Hybrid1} \cite{Hybrid2}.

The contribution of the current study is threefold:

\begin{enumerate}
 
 \item We introduce the use of deep learning for on-line correction of motion related errors in abdominal MR-thermometry. The existing multi-baseline strategy is extended by a convolutional neural network (CNN) which learns the apparent temperature perturbation from images acquired during the preparative learning stage. The input of the designed CNN is the current magnitude image and as a consequence no surrogate of motion-state is needed. During the hyperthermia procedure, the most recent magnitude image is used as an input for the pre-built CNN-model in order to generate a correction for the most recent temperature map.
 
 \item Frequently, inherent computational costs are a major difficulty when dealing with deep learning in applications requiring adaptive / on-the-fly training. In order to mitigate this problem for clinical applications of MR-thermometry, a fine-tuning strategy is proposed to accelerate calculations during the preparative learning stage. Moreover, to meet computational requirements for real-time MR-thermometry, which requires that all calculations have to be completed within the interval between two successive image acquisitions, we propose to process all temperature images in a sliding temporal window within one single CNN-model call. 
 
 \item The ability of the proposed approach to remove thermometry artifacts is demonstrated for dynamic MRI datasets of the the liver of 12 healthy volunteers in absence of temperature changes. We demonstrate that the amount of learning images and the CNN training time can be optimized to the point that clinical thermotherapy interventions are feasible. Thermometric precision and accuracy is demonstrated with a heating experiment performed on a porcine liver using high intensity focused ultrasound (HIFU) \cite{hifu2}. The proposed method is compared to the two most frequently employed multi-baseline strategies in terms of temperature precision, without penalty in accuracy.
 
\end{enumerate}

\section{Materials and Methods}

\subsection{Method overview}

\begin{figure}[h!]
\begin{minipage}[b]{\linewidth}
\centering
\centerline{\includegraphics[width=\linewidth]{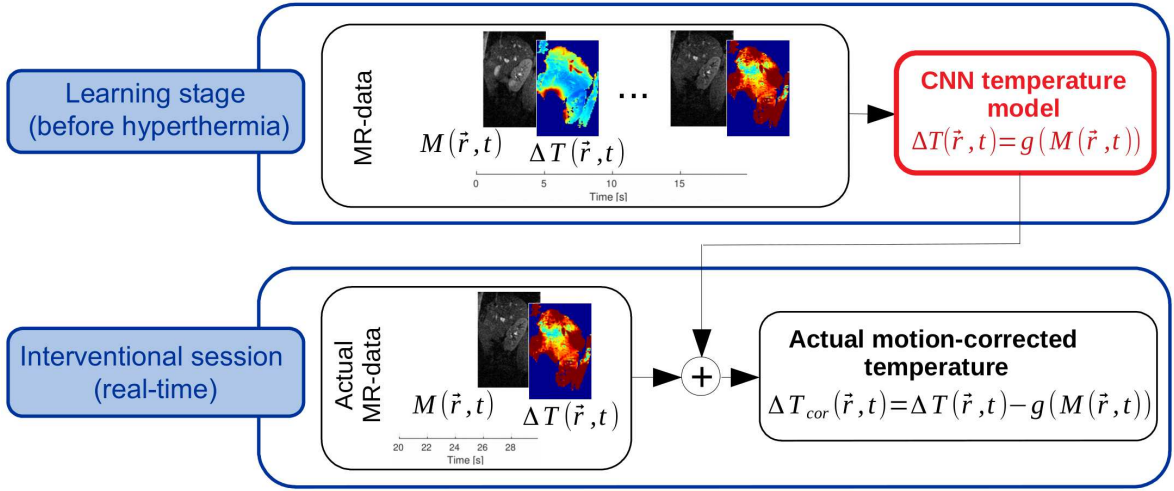}}
\end{minipage}
\caption{Illustration of the proposed CNN correction scheme. In this approach only MR-images are required and no motion surrogate(s)/sensors are needed. MR-training data are acquired prior to hyperthermia in a training phase. Subsequently, the motion-artifacted thermal maps and the magnitude images are combined to establish a CNN-based correction model of the temperature (red block). During the hyperthermia procedure, the most recent magnitude image is used as an input for the pre-built CNN-model in order to generate a correction map (noted $g(M)$) for the current temperature map.}
\label{fig:CNN_scheme}
\end{figure}

The proposed method is detailed in figure \ref{fig:CNN_scheme}: thermal maps with motion related temperature artifacts and magnitude images (noted $M$) are combined to establish prior to hyperthermia a CNN-based temperature correction model (noted $g$). During hyperthermia, the incoming magnitude image is used as an input for the pre-built CNN-model to generate in real time a temperture correction for the current temperature map. Differences with existing multi-baseline approaches (figure \ref{fig:MB_scheme}) are: (i) the input of the correction model is the most recent magnitude image, which eliminates the need for surrogates/sensors ; (ii) the fitted data is the apparent (artifacted) temperature ; (iii) the model is a CNN. The benefit of each of these aspects is discussed later in the manuscript.

\subsection{Datasets}

Dynamic MR-imaging was performed on a Philips Achieva 1.5 T (Philips Healthcare, Best, The Netherlands) under real-time conditions. The method was evaluated in 2D and the effect of through plane motion was reduced by setting the imaging plane direction parallel to the principal axis of the organ displacement.

\subsubsection{Volunteer study.}

An imaging frame rate of 10 Hz was maintained during 5 minutes on the abdomen of 12 healthy human volunteers under free-breathing conditions. The MR-protocol was composed of a learning step of 20 s dedicated to the acquisition of the training data, followed by 4 min-40 s devoted to mimic an interventional procedure. The MR-sequence was a single-shot gradient recalled echo-planar with the following parameters: one coronal slice, repetition time ($TR$)=100 ms, echo time ($TE$)=26 ms, bandwidth in readout direction=2085 Hz, flip angle=35$^\circ$, field of view ($FOV$)=$256\times 168$ mm$^2$, slice thickness=6 mm, matrix=$128\times 84$, using a four element phased array body coil. The volunteer studies depicted various SNR conditions: over the twelve volunteers, the SNR was evaluated as $7 \pm 3$ (min=4, max=14) in the liver.

\subsubsection{\textit{In vivo} heating study in a porcine liver.}

MRI guided HIFU was performed \textit{in vivo} in the liver of a pig under general anesthesia and artificial breathing. The MR sequence employed the following parameters: single-shot, gradient recalled, echo-planar imaging, 1000 dynamic sagittal images, five slice, $TR$=250 ms, $TE$=33 ms, flip angle=40$^{\circ}$, in-plane $FOV$=$370 \times 162$ mm$^2$, voxel size=$2.89 \times 2.89 \times 7$ mm$^3$ using the integrated three elements phased array coil of the HIFU system. A MR compatible HIFU ablation system (Sonalleve, Profound Medical, Helskinki, Finland) composed of a table top containing a 256 elements HIFU transducer, integrated in the 1.5 T Achieva-Intera MRI was used to perform a temperature elevation. The transducer radius and aperture were 120 mm and 126 mm, respectively, creating an ellipsoid focal point ($1 \times 1 \times 7$ mm$^3$). The animals were placed in the prone position so that the liver was accessible through an unobstructed beam-path directly below the rib-cage. MR-guided hyperthermia was performed for a duration of 4 minutes on the liver with HIFU power of 160 W during 100 s. All animal studies were performed under an approved Animal Care and Use protocol.

\subsection{Proposed CNN-based correction}

\subsubsection{Learning motion-related errors in MR-thermometry.}

Motion-related errors in MR-thermometry were learned during a preparative learning stage performed before hyperthermia. This step is based on a training set of $N$ dynamically acquired data (each dynamic data is composed by the magnitude, the phase and the apparent temperature map calculated with Eq. (\ref{eq:PRF})). The motion cycle has to be sampled with a sufficient density in order to avoid discretization errors. With a sufficient imaging frame rate of 5-10 Hz and a respiration frequency of 3-6 seconds this pre-treatment step can be completed in a relatively short duration of 15-20 seconds. For the volunteer study, we tested various (reconstructed) imaging frame rates --- ranking from 1 Hz (\emph{i.e.}, $N=20$) to 10 Hz (\emph{i.e.}, $N=200$) --- to train our CNN-model. For the \textit{in vivo} heating study, we used $N=200$.

\subsubsection{Preprocessing of input images.}

During both learning and hyperthermia (thermotherapy) stages, all incoming images (\emph{i,e.} anatomical and temperature images) were preprocessed on-the-fly as follows. First, anatomical (magnitude) image intensities were normalized (z-scoring) using the mean and standard deviation within the complete image field-of-view. Second, thermal maps were registered onto a common reference position in order to allow kinetic analysis. Note that this registration step also allows for cumulative thermal dose calculations, which may be beneficial for on-line assessment of the therapy endpoint \cite{necrosis} \cite{TD_eval}. In the scope of this manuscript, we registered all incoming phase images using motion estimates of a real-time image optical-flow (OF) algorithm applied to magnitude images. Additional details about the employed image registration algorithm can be found in \cite{Zachiu15}. Temperature calculations were performed using Eq. (\ref{eq:PRF}) applied to the registered phase maps.

\subsubsection{Implemented deep neural network model.} 

\begin{figure}[h!]
\begin{minipage}[b]{\linewidth}
\centering
\centerline{\includegraphics[width=\linewidth]{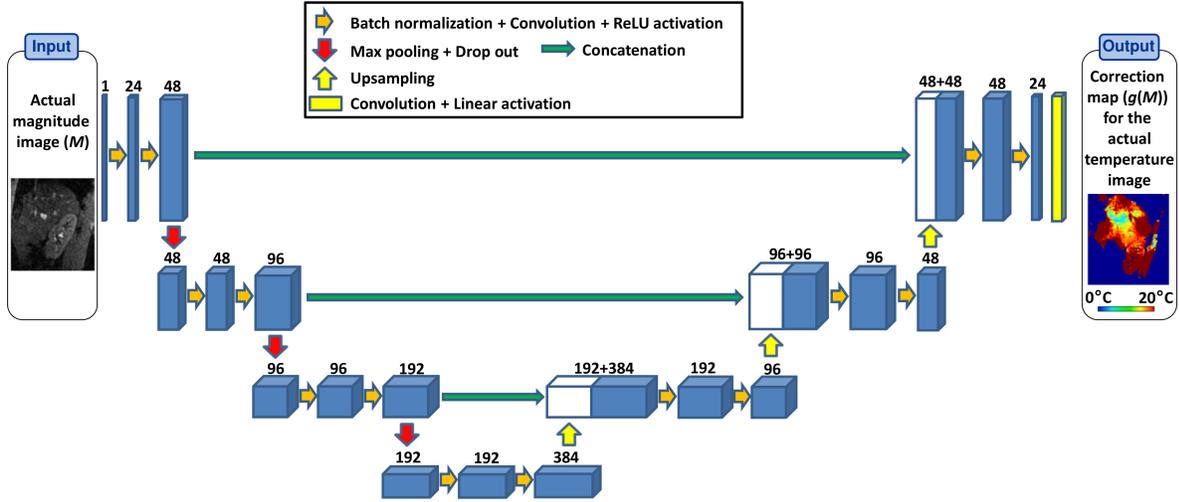}}
\end{minipage}
\caption{Architecture of the deep neural network proposed for learning motion-related errors in MR-thermometry before heating (red block in figure \ref{fig:CNN_scheme}). The most recent magnitude image (\emph{i.e.}, $M$) is used as a single input channel. The CNN-model constructs a correction map (noted $g(M)$) for the current temperature map. Each block of the CNN is composed of batch normalization, convolution and ReLU activation. The number of $3 \times 3$ filters is indicated on the top of each block.}
\label{fig:CNN_architecture}
\end{figure}

Figure \ref{fig:CNN_architecture} describes the architecture of the proposed deep neural network model (noted $g$) designed to learn motion-related artifacts in MR-thermometry prior to hyperthermia (red block in figure \ref{fig:CNN_scheme}). The input of the model is a magnitude image $M$ and the output is a correction (noted $g(M)$) for the corresponding temperature image (\emph{i.e.}, $\Delta T$). Note that potential mis-corrected phase wraps in Eq. (\ref{eq:PRF}) may have a direct negative impact on the CNN-model optimization process. To mitigate this drawback, temperature maps used for training were filtered using a median filter (kernel $5 \times 5$). We used a convolutional encoder with 3 layers per resolution level, using a basis of 24 filters of $3 \times 3$ (\emph{i.e.}, 24 filters for the first layer, 48 for the second and so on). We empirically optimized this setting for reduced memory consumption without impacting performance. Each block was composed of batch normalization, convolution and ReLU activation. We employed the following parameters: batch size = 1, optimizer = Adam with default parameters, epoch = 100, loss = Mean Square Error (MSE) and dropout = 0.5 after each block. We used one single input channel (\emph{i.e.}, the actual magnitude image).
The implemented CNN-model is detailed in the supplementary material of the manuscript (see Table \ref{table:model}). The output shape and the number of parameters involved in each layer of the CNN-model are given.

\subsubsection{Proposed fine-tuning strategy.}

CNN optimization is a complex iterative process which is inherently time consuming and depends on initialization. To reduce these two issues and to meet clinical constraints of a therapeutic workflow, we propose a fine-tuning strategy. To this end, a pre-built CNN model (\emph{i.e.,} a CNN-model trained on several data sets) --- is loaded and used as a starting point for the actual model optimization (\emph{i.e.,} the red block in figure \ref{fig:CNN_scheme}).

\subsubsection{On-line CNN-correction of temperature maps.}
\label{sssec:CNNTemp}

At this point we have a model $g$ designed to predict the actual temperature perturbation $g(M)$ given the current anatomical image $M$. The motion compensated temperature image at instant $t$ can be obtained as follows (see figure \ref{fig:CNN_scheme}):

\begin{equation}
 \Delta T_{cor}(\vec{r},t) = \Delta T( \vec{r}, t ) - g\left( M ( \vec{r}, t ) \right)
\end{equation}

\subsubsection{Correction of time-persistent offsets.}
\label{sssec:bias}

Once corrected according to section \ref{sssec:CNNTemp}, the temperature in a voxel at location $\vec{r}$ is prone to a time dependent offset arising from the presence of noise in the baseline phase image (\emph{i.e.}, $\varphi (\vec{r},t_0)$ in Eq. (\ref{eq:PRF})). To compensate for this offset, we assumed that the temperature change has to be identically equal to 0 during the learning stage (no hyperthermia). Practically, a pre-built temperature time average map --- based on data acquired before heating --- was subtracted from the actual motion compensated temperature image $\Delta T_{cor}(\vec{r},t)$.

\subsection{Implementation details}
\label{sssec:temp_win}

We evaluated the computational burden of our proposed method using an Intel Xeon E5-2683 2.4 GHz (2 Hexadeca-core) with 256 GB of RAM equipped by a GPU Nvidia Tesla V100. Our implementation was performed using Tensorflow 1.4 and Keras 2.2.4.

Each of the 13 tested data sets --- 12 volunteer data sets + 1 \textit{in vivo} heating data set --- were processed without fine-tuning (each data set was processed independently of each other) and with fine-tuning (leave-one-out strategy).

To reduce the computation time during the interventional procedure, all data in a sliding temporal window (\emph{i.e.}, a pack of most recent consecutive dynamically acquired images. We denote by $\delta$ the number of dynamic images) were processed simultaneously within one single CNN-model call. In the scope of this study, we tested $\delta$-values in the following set: $\{1,2,16\}$.

The usefulness of the two above-mentioned implementation strategies is analyzed in the discussion section.

\subsection{Validation framework}

\subsubsection{Implemented competitive approaches.}

The performance of two existing multi-baseline solutions --- referred to as ``look-up-table'' approach (or LUT) \cite{gating2} and ``linear model'' approach (or LM) \cite{pipeline} throughout the rest of the manuscript --- were also evaluated on the same data sets. In the learning stage of both LUT and LM, MR-images (magnitude and phase) were included over several respiratory cycles (identical number of training data $N$ were employed for both LUT, LM and the proposed CNN method). 
The position in the respiration cycle (\emph{i.e.,} the motion surrogate denoted by $S(t)$ in figure \ref{fig:MB_scheme}) was monitored using a Principal Component Analysis (PCA) applied to the above-mentioned OF-motion estimates, as described in \cite{PCA_MRgHIFU}. The baseline phase image needed for the calculation of the actual temperatures maps with Eq. (\ref{eq:PRF}) was calculated in the following two ways:

\paragraph{Look-up-table approach (LUT):}

Each collected baseline phase image during the learning stage was indexed in a look-up-table according to its estimated position within the breathing cycle given by $S(t)$. During the intervention stage, the baseline phase image in Eq. (\ref{eq:PRF}) was a linear interpolation between the closest two reference phase images allowed for reconstructing a baseline phase image for the current position in the respiration cycle.

\paragraph{Linear model approach (LM):}

The overall phase variation (denoted by $f(S(t))$ in figure \ref{fig:MB_scheme}) was approximated by linear phase changes of the motion surrogate $S(t)$ on a voxel-by-voxel basis as described in \cite{PCA_MRgHIFU}.

\subsubsection{Statistical analysis.} 

In the volunteer study, it was assumed that the temperature change has to be identically equal to 0 during the testing session (no hyperthermia was performed). 

First, the temperature precision was evaluated for each volunteer by computing on a voxel-by-voxel basis the temporal temperature standard deviation (noted SD) within a manually defined mask (noted $\Gamma$, $\Gamma \subset \Omega$) encompassing the liver and over the duration of the interventional session (\emph{i.e.,} from the starting instant $t_s=20$ s to the final instant $t_f=$5 min):

\begin{equation}
 \SD(\vec{r}) = \sigma \left( \Delta T( \vec{r}, t ) \right)\hspace{1.5cm} t \in [t_s, t_f], \hspace{0.5cm} \vec{r} \in \Gamma
\end{equation}

Second, the temperature accuracy was evaluated for each volunteer by computing on a voxel-by-voxel basis the mean absolute temperature error (noted MAE) within $\Gamma$ and over the interventional step:

\begin{equation}
 \MAE(\vec{r}) = \left| \frac{1}{t_f-t_s} \int _{t=t_s} ^{t_f} \Delta T( \vec{r}, t ) \mathrm{d}t \right| \hspace{1.28cm} \vec{r} \in \Gamma
\end{equation}

The same analysis was performed for the \textit{in vivo} heating study to assess the thermometry precision and the accuracy outside the heated region.

For the volunteer study, a paired Wilcoxon test was carried out in order to study whether SD and MAE differences are statistically significant between LUT-, LM- and CNN-corrected data sets. A significance threshold of $p=0.025$ was used. The power of the statistical analysis has been carried out, as described in \cite{power_analysis}.

\section{Results}

\subsection{Volunteer study}

\begin{figure}[h!]

\begin{minipage}[b]{0.22\linewidth}
\centering
\vspace{3cm}
\centerline{Magnitude}\medskip
\centerline{\includegraphics[trim={0cm 0cm 0cm 0cm},clip,height=4cm]{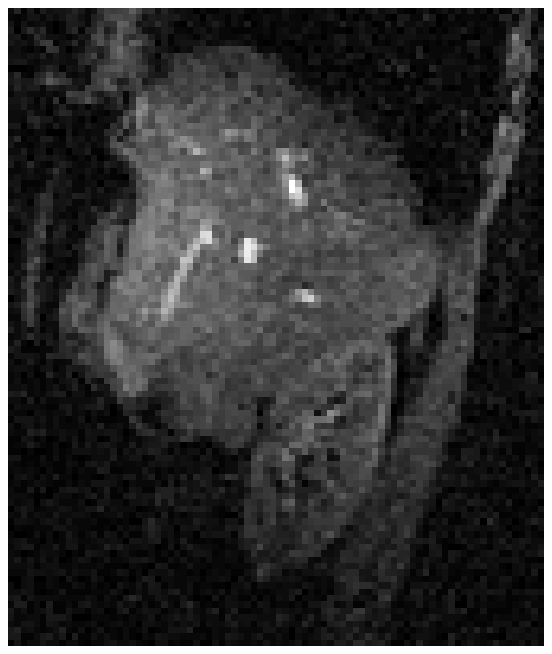}}
\centerline{(a)}\medskip
\vspace{-3cm}
\end{minipage}
\begin{minipage}[b]{0.22\linewidth}
\centerline{SD-map (LUT)}\medskip
\centerline{\includegraphics[trim={0cm 0cm 0cm 0cm},clip,height=4cm]{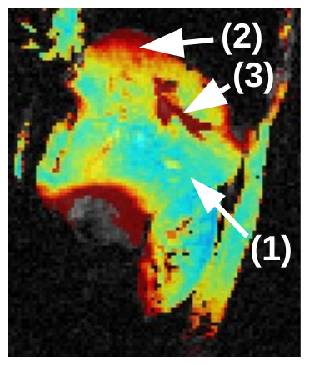}}
\centerline{(b)}\medskip
\end{minipage}
\begin{minipage}[b]{0.22\linewidth}
\centerline{SD-map (LM)}\medskip
\centerline{\includegraphics[trim={0cm 0cm 0cm 0cm},clip,height=4cm]{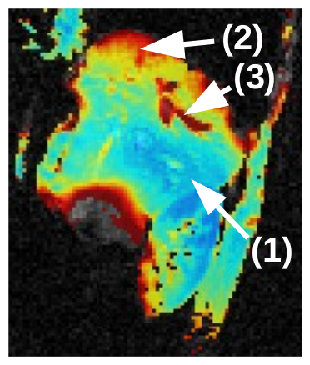}}
\centerline{(c)}\medskip
\end{minipage}
\begin{minipage}[b]{0.22\linewidth}
\centerline{SD-map (CNN)}\medskip
\centerline{\includegraphics[trim={0cm 0cm 0cm 0cm},clip,height=4cm]{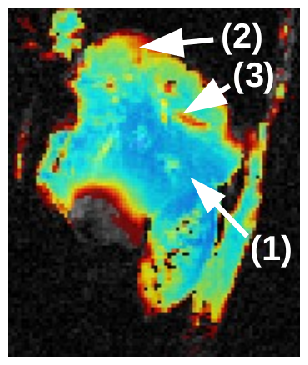}}
\centerline{(d)}\medskip
\end{minipage}
\begin{minipage}[b]{0.07\linewidth}
\vspace{1.9cm}
\centerline{\includegraphics[trim={0cm 0cm 0cm 0cm},clip,height=4cm]{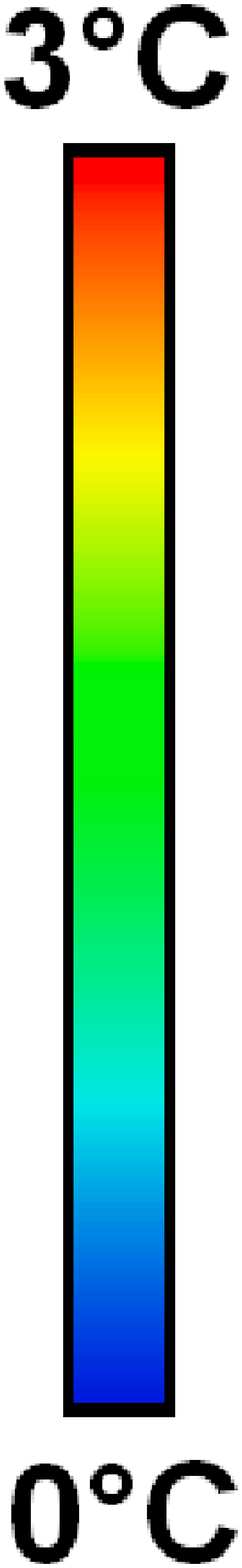}}
\vspace{-1.9cm}
\end{minipage}

\begin{minipage}[b]{0.22\linewidth}
\centering
\centerline{ }\medskip
\end{minipage}
\begin{minipage}[b]{0.22\linewidth}
\centerline{MAE-map (LUT)}\medskip
\centerline{\includegraphics[trim={0cm 0cm 0cm 0cm},clip,height=4cm]{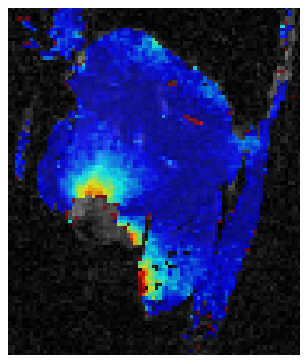}}
\centerline{(e)}\medskip
\end{minipage}
\begin{minipage}[b]{0.22\linewidth}
\centerline{MAE-map (LM)}\medskip
\centerline{\includegraphics[trim={0cm 0cm 0cm 0cm},clip,height=4cm]{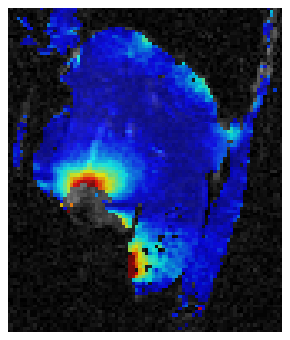}}
\centerline{(f)}\medskip
\end{minipage}
\begin{minipage}[b]{0.22\linewidth}
\centerline{MAE-map (CNN)}\medskip
\centerline{\includegraphics[trim={0cm 0cm 0cm 0cm},clip,height=4cm]{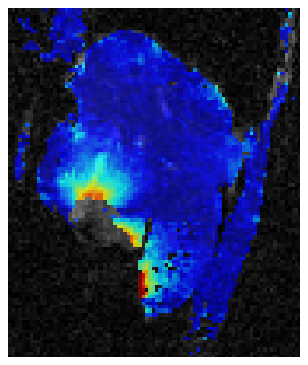}}
\centerline{(g)}\medskip
\end{minipage}
\begin{minipage}[b]{0.07\linewidth}
\hspace{1cm}
\end{minipage}
\caption{Typical temperature stability maps obtained in the abdomen of volunteer $\#2$ using two existing multi-baseline approaches (\emph{i.e.,} LUT and LM) and using the CNN approach: (a) anatomic image, (upper row) the temperature standard deviation map obtained with the LUT (b), the LM (c) and the CNN method (d), (lower row) the temperature mean absolute error map obtained with LUT (e), LM (f) and CNN (g). $N=200$ images were used for training (an imaging frame rate of 10 hz was maintained during 20 seconds)}
\label{fig:Thermo_map}
\end{figure}

Figure \ref{fig:Thermo_map} shows an example of MR-thermometry results obtained in one volunteer of the examined group (volunteer $\#2$). The leftmost image (\ref{fig:Thermo_map}a) depicts the anatomy. The temperature precision (resp. accuracy) is reported in the upper row (resp. bottom row) for each tested correction solution. Thermometry artifacts caused by motion-related susceptibility variations were compensated using LUT (first column), LM (second column) and CNN (third column). The temperature precision is visually better in the major part of the liver using LM method as compared to LUT (see arrow $\#1$ in \ref{fig:Thermo_map}b and \ref{fig:Thermo_map}c). The best precision is however observable using the CNN method. It can also be noticed that large susceptibility artifacts render the temperature correction difficult in the upper part of the liver (see arrow $\#2$ in \ref{fig:Thermo_map}b, \ref{fig:Thermo_map}c and \ref{fig:Thermo_map}d) and in the vicinity of hepatic arteries (see arrow $\#3$ in \ref{fig:Thermo_map}b, \ref{fig:Thermo_map}c and \ref{fig:Thermo_map}d). In these regions, an improvement of the thermometry precision by up to 2 $^{\circ}$C could be obtained using CNN as compared to both LUT and LM. This precision was achieved without negative impact on the accuracy, from a visual point of view, as shown in \ref{fig:Thermo_map}e-g.

\begin{figure}[h!]
\begin{minipage}[b]{0.18\linewidth}
\centering
\vspace{3cm}
\centerline{\includegraphics[trim={0cm 0cm 0cm 0cm},clip,height=5cm]{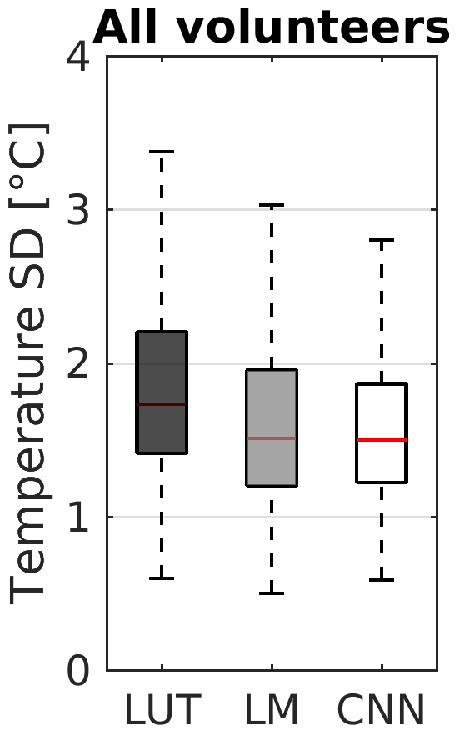}}
\centerline{(a)}\medskip
\vspace{-3cm}
\end{minipage}
\begin{minipage}[b]{0.13\linewidth}
\centering
\centerline{\includegraphics[trim={0cm 0cm 0cm 0cm},clip,height=3.8cm]{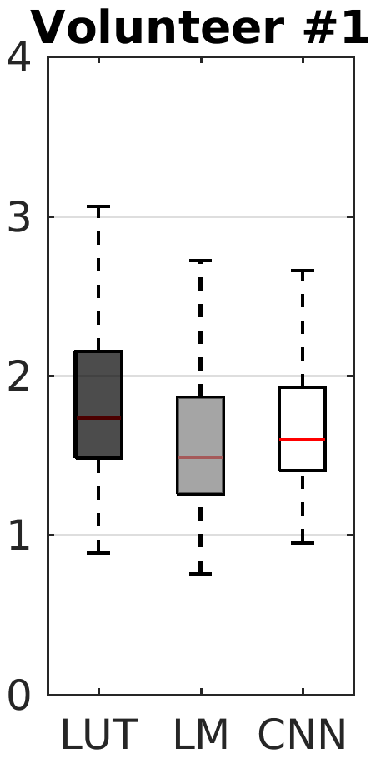}}
\end{minipage}
\begin{minipage}[b]{0.13\linewidth}
\centerline{\includegraphics[trim={0cm 0cm 0cm 0cm},clip,height=3.8cm]{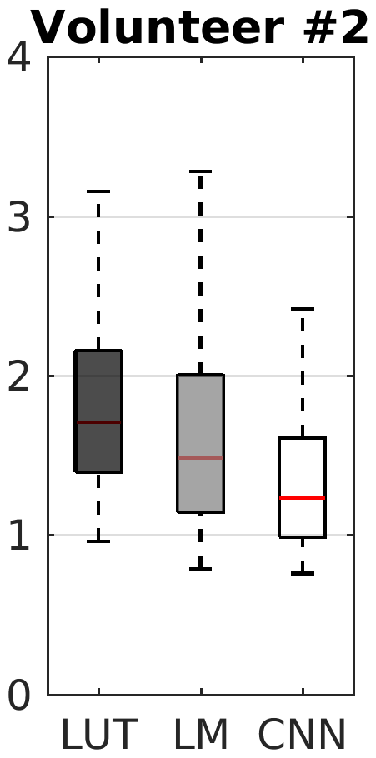}}
\end{minipage}
\begin{minipage}[b]{0.13\linewidth}
\centerline{\includegraphics[trim={0cm 0cm 0cm 0cm},clip,height=3.8cm]{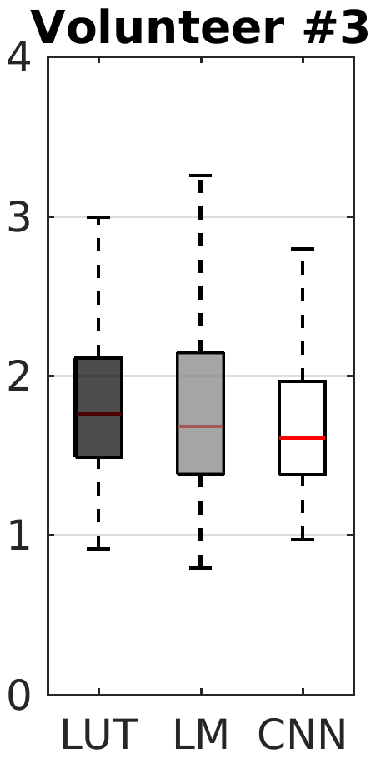}}
\end{minipage}
\begin{minipage}[b]{0.13\linewidth}
\centering
\centerline{\includegraphics[trim={0cm 0cm 0cm 0cm},clip,height=3.8cm]{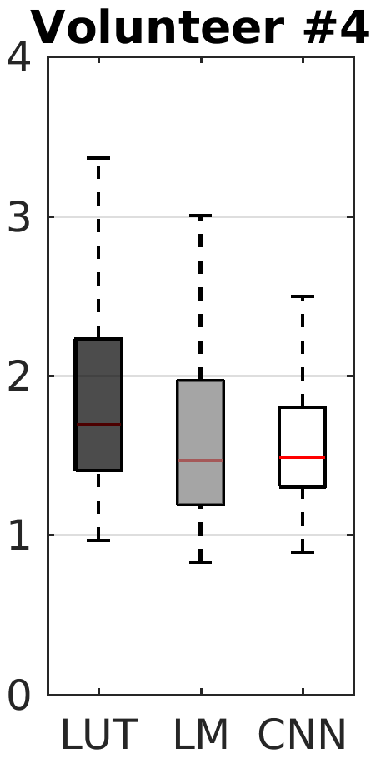}}
\end{minipage}
\begin{minipage}[b]{0.13\linewidth}
\centerline{\includegraphics[trim={0cm 0cm 0cm 0cm},clip,height=3.8cm]{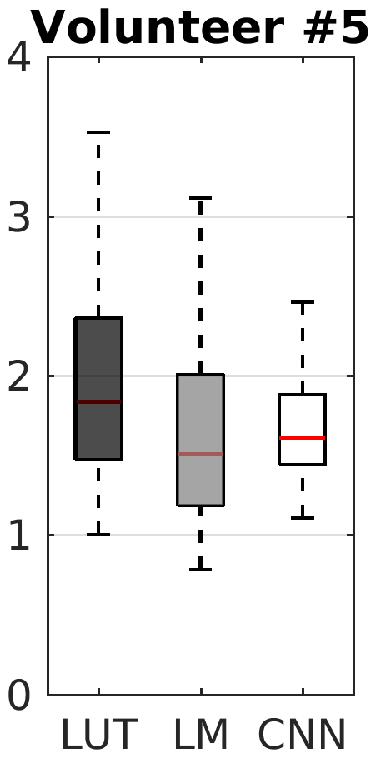}}
\end{minipage}
\begin{minipage}[b]{0.13\linewidth}
\centerline{\includegraphics[trim={0cm 0cm 0cm 0cm},clip,height=3.8cm]{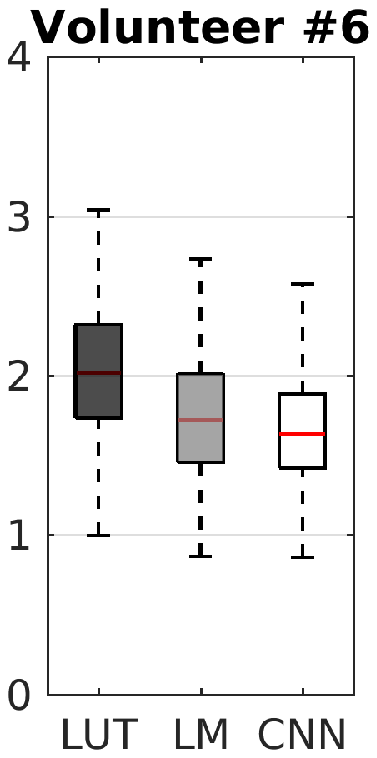}}
\end{minipage}

\begin{minipage}[b]{0.18\linewidth}
\centering
\centerline{ }\medskip
\end{minipage}
\begin{minipage}[b]{0.13\linewidth}
\centering
\centerline{\includegraphics[trim={0cm 0cm 0cm 0cm},clip,height=3.8cm]{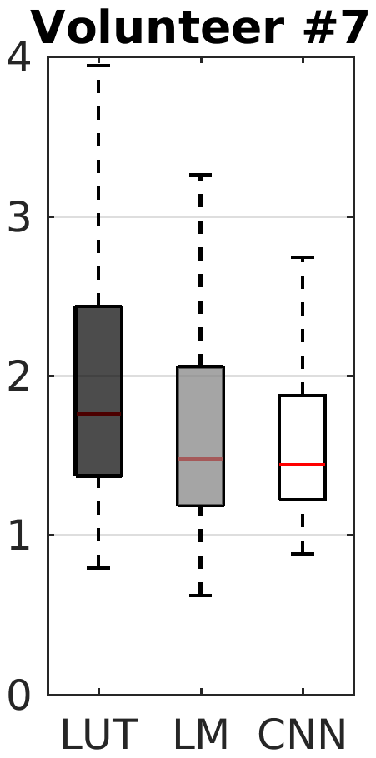}}
\centerline{ }\medskip
\end{minipage}
\begin{minipage}[b]{0.13\linewidth}
\centerline{\includegraphics[trim={0cm 0cm 0cm 0cm},clip,height=3.8cm]{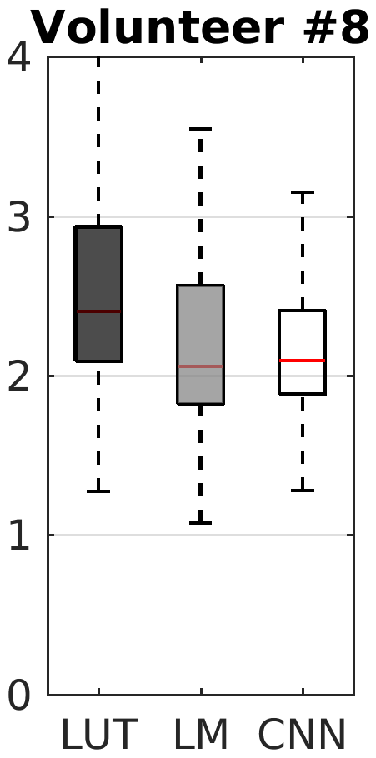}}
\centerline{ }\medskip
\end{minipage}
\begin{minipage}[b]{0.13\linewidth}
\centerline{\includegraphics[trim={0cm 0cm 0cm 0cm},clip,height=3.8cm]{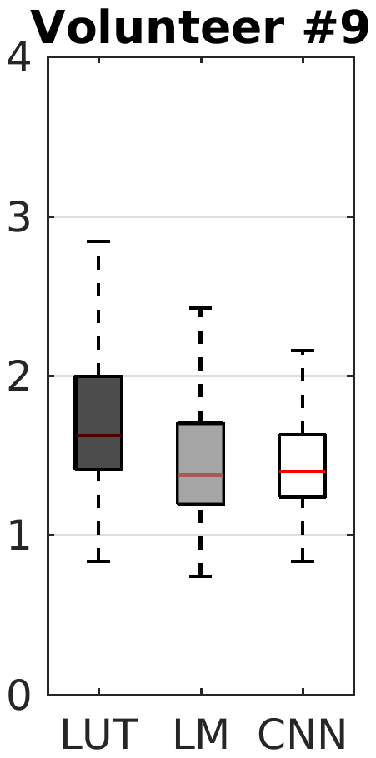}}
\centerline{\hspace{1cm}(b)}\medskip
\end{minipage}
\begin{minipage}[b]{0.13\linewidth}
\centering
\centerline{\includegraphics[trim={0cm 0cm 0cm 0cm},clip,height=3.8cm]{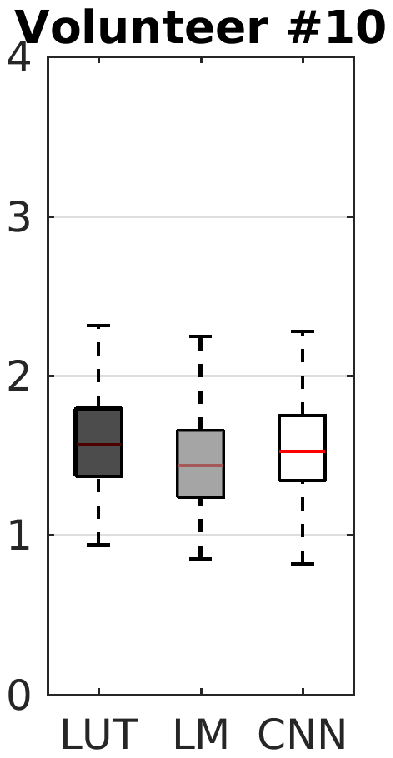}}
\centerline{ }\medskip
\end{minipage}
\begin{minipage}[b]{0.13\linewidth}
\centerline{\includegraphics[trim={0cm 0cm 0cm 0cm},clip,height=3.8cm]{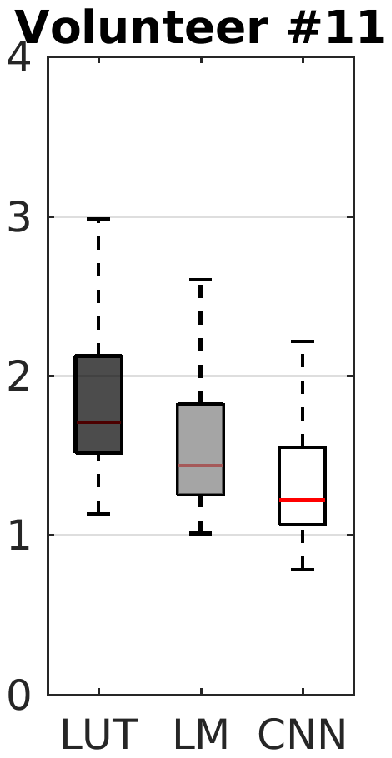}}
\centerline{ }\medskip
\end{minipage}
\begin{minipage}[b]{0.13\linewidth}
\centerline{\includegraphics[trim={0cm 0cm 0cm 0cm},clip,height=3.8cm]{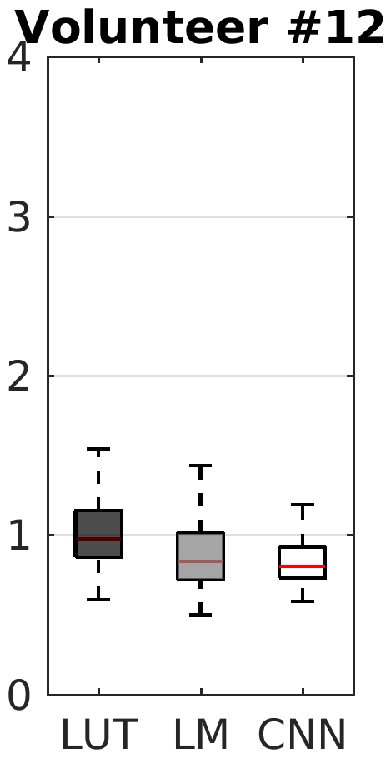}}
\centerline{ }\medskip
\end{minipage}
\caption{Comparison of temperature precision obtained using LUT, LM and CNN in the liver of free-breathing healthy volunteers during 4 minutes and 40 seconds of MR-thermometry. Similar to figure \ref {fig:Thermo_map}, $N=200$ images were used for training. Box-and-whisker plots of the temporal temperature standard deviation are shown using LUT (dark gray box), LM (light gray box) and CNN (white box): (a) group analysis over the 12 volunteers, (b) volunteer-wise analysis. The median is shown by the central mark, the first and the third quartiles are reported by the edges of the box, the whiskers extend to the most extreme time points that are not considered as outliers.}
\label{fig:SD_liver}
\end{figure}

\begin{figure}[h!]
\begin{minipage}[b]{0.18\linewidth}
\centering
\vspace{3cm}
\centerline{\includegraphics[trim={0cm 0cm 0cm 0cm},clip,height=5cm]{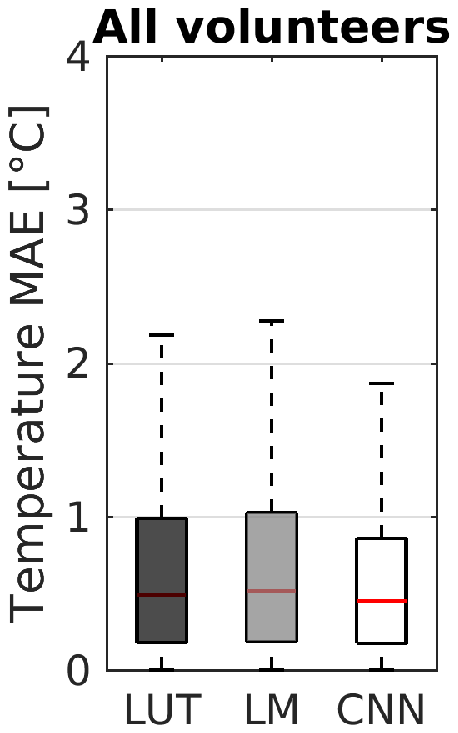}}
\centerline{(a)}\medskip
\vspace{-3cm}
\end{minipage}
\begin{minipage}[b]{0.13\linewidth}
\centering
\centerline{\includegraphics[trim={0cm 0cm 0cm 0cm},clip,height=3.8cm]{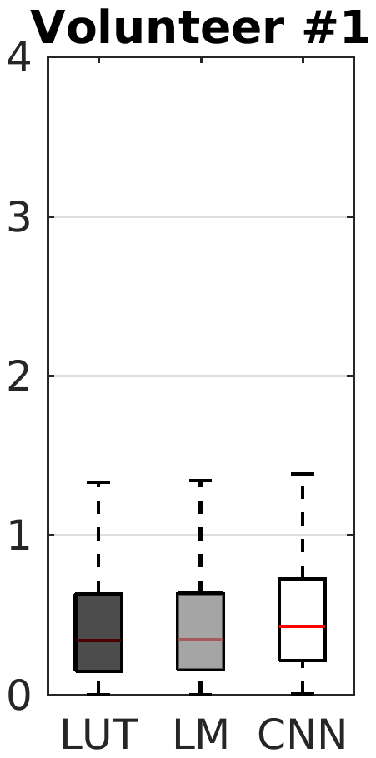}}
\end{minipage}
\begin{minipage}[b]{0.13\linewidth}
\centerline{\includegraphics[trim={0cm 0cm 0cm 0cm},clip,height=3.8cm]{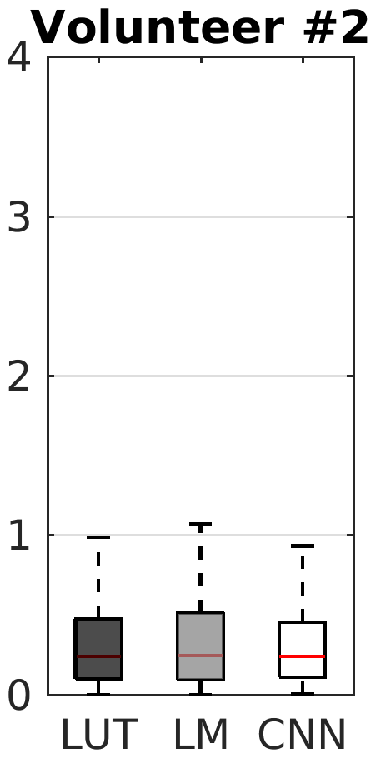}}
\end{minipage}
\begin{minipage}[b]{0.13\linewidth}
\centerline{\includegraphics[trim={0cm 0cm 0cm 0cm},clip,height=3.8cm]{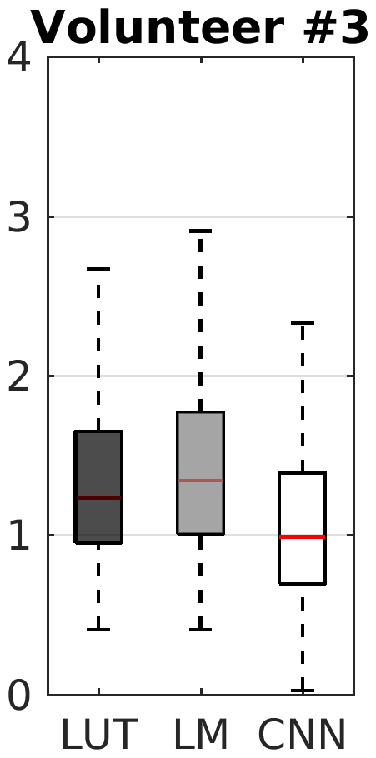}}
\end{minipage}
\begin{minipage}[b]{0.13\linewidth}
\centering
\centerline{\includegraphics[trim={0cm 0cm 0cm 0cm},clip,height=3.8cm]{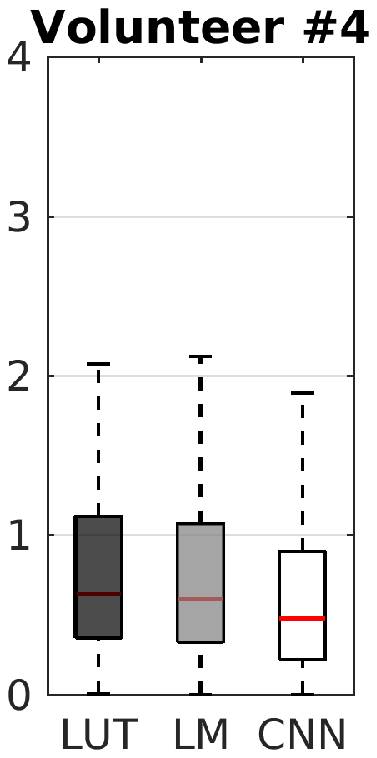}}
\end{minipage}
\begin{minipage}[b]{0.13\linewidth}
\centerline{\includegraphics[trim={0cm 0cm 0cm 0cm},clip,height=3.8cm]{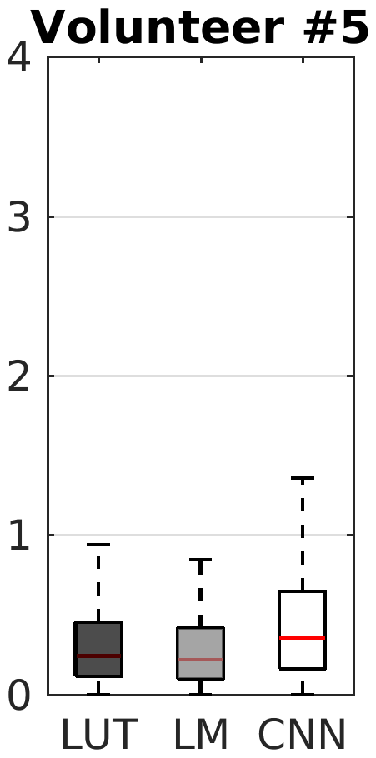}}
\end{minipage}
\begin{minipage}[b]{0.13\linewidth}
\centerline{\includegraphics[trim={0cm 0cm 0cm 0cm},clip,height=3.8cm]{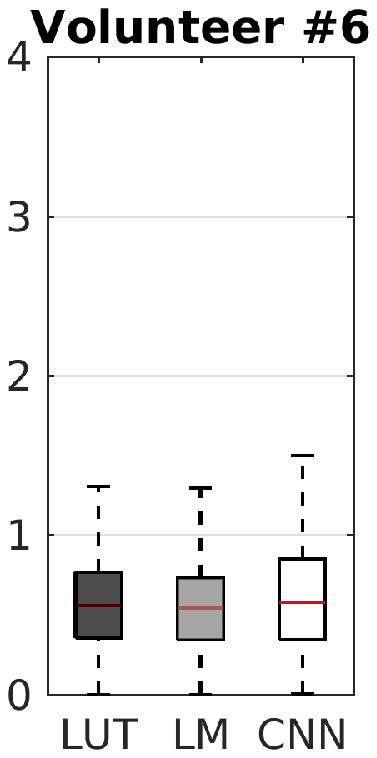}}
\end{minipage}

\begin{minipage}[b]{0.18\linewidth}
\centering
\centerline{ }\medskip
\end{minipage}
\begin{minipage}[b]{0.13\linewidth}
\centering
\centerline{\includegraphics[trim={0cm 0cm 0cm 0cm},clip,height=3.8cm]{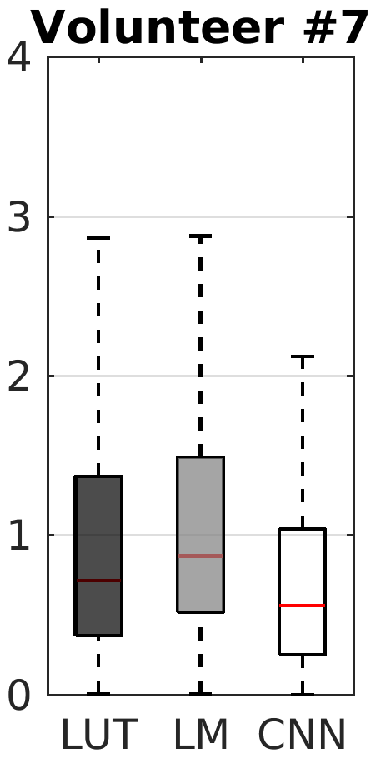}}
\centerline{ }\medskip
\end{minipage}
\begin{minipage}[b]{0.13\linewidth}
\centerline{\includegraphics[trim={0cm 0cm 0cm 0cm},clip,height=3.8cm]{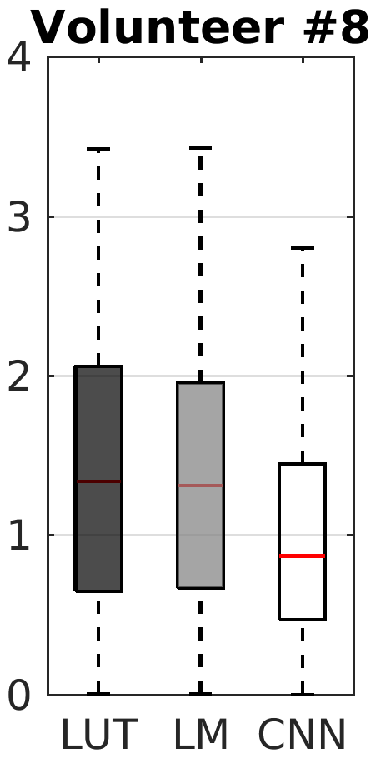}}
\centerline{ }\medskip
\end{minipage}
\begin{minipage}[b]{0.13\linewidth}
\centerline{\includegraphics[trim={0cm 0cm 0cm 0cm},clip,height=3.8cm]{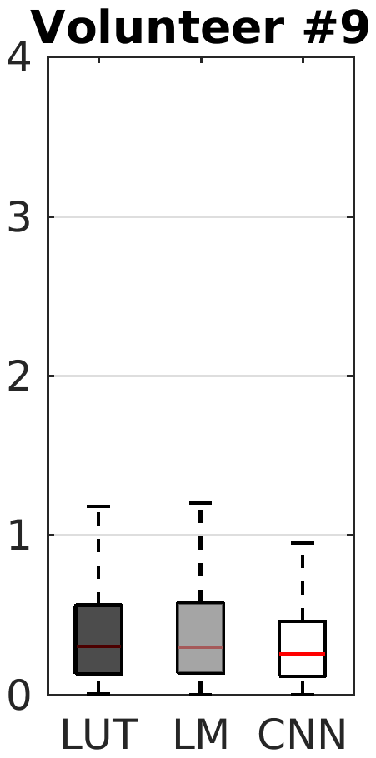}}
\centerline{\hspace{1cm}(b)}\medskip
\end{minipage}
\begin{minipage}[b]{0.13\linewidth}
\centering
\centerline{\includegraphics[trim={0cm 0cm 0cm 0cm},clip,height=3.8cm]{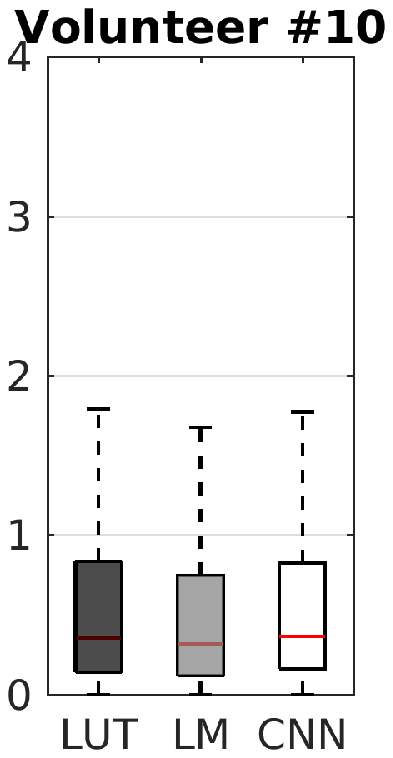}}
\centerline{ }\medskip
\end{minipage}
\begin{minipage}[b]{0.13\linewidth}
\centerline{\includegraphics[trim={0cm 0cm 0cm 0cm},clip,height=3.8cm]{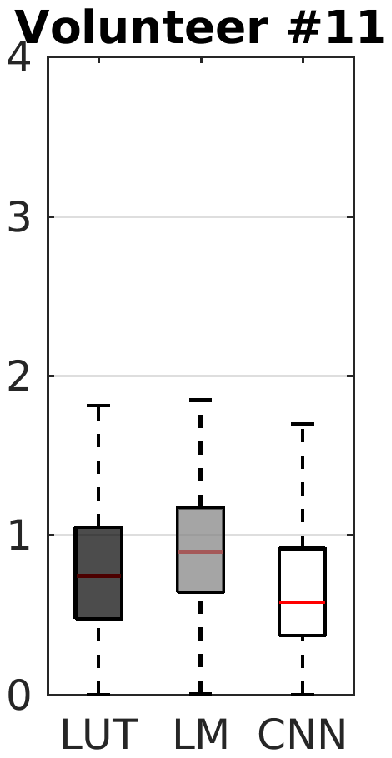}}
\centerline{ }\medskip
\end{minipage}
\begin{minipage}[b]{0.13\linewidth}
\centerline{\includegraphics[trim={0cm 0cm 0cm 0cm},clip,height=3.8cm]{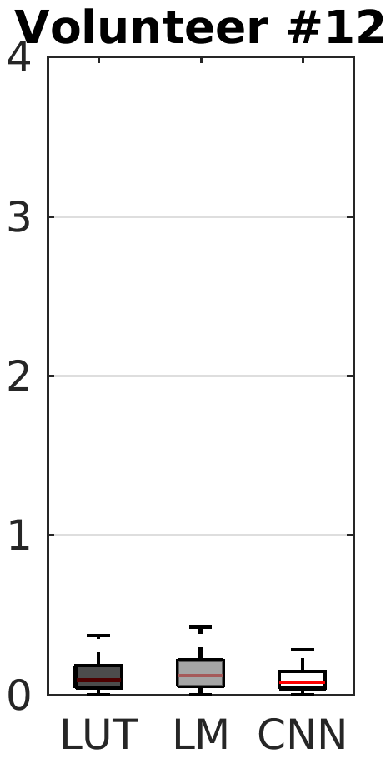}}
\centerline{ }\medskip
\end{minipage}
\caption{Comparison of temperature accuracy obtained using LUT, LM and CNN in the liver of free-breathing healthy volunteers during 4 minutes and 40 seconds of MR-thermometry. Similar to figures \ref {fig:Thermo_map} and \ref{fig:MAE_liver}, $N=200$ images were used for training. Box-and-whisker plots of the temporal temperature mean absolute error are shown using LUT (dark gray box), LM (light gray box) and CNN (white box): (a) group analysis over the 12 volunteers, (b) volunteer-wise analysis.}
\label{fig:MAE_liver}
\end{figure}

Temperature maps from the individual volunteers were pooled in order to obtain a group set containing the temperature precision (resp. accuracy) from all volunteers. The distribution of the temperature precision (resp. accuracy) for the group set is reported in figure \ref{fig:SD_liver}a (resp. \ref{fig:MAE_liver}a) using LUT, LM and CNN. The distribution of temperature precision (resp. accuracy) is also detailed for each volunteer in \ref{fig:SD_liver}b (resp. \ref{fig:MAE_liver}b). The paired Wilcoxon test showed that the temperature precision was significantly better using LM as compared to LUT (p$<$0.001/statistical power=1). Furthermore, the temperature precision was significantly better using CNN as compared to LUT and LM (p$<$0.001/statistical power=1). Besides, the temperature accuracy was significantly better using CNN as compared to LM (p=0.008/statistical power=1).

\begin{figure}[h!]
\begin{minipage}[b]{0.49\linewidth}
\centerline{\large{Temperature precision}}\medskip
\vspace{-0.2cm}
\centerline{\includegraphics[trim={0cm 0cm 0cm 0cm},clip,height=5.5cm]{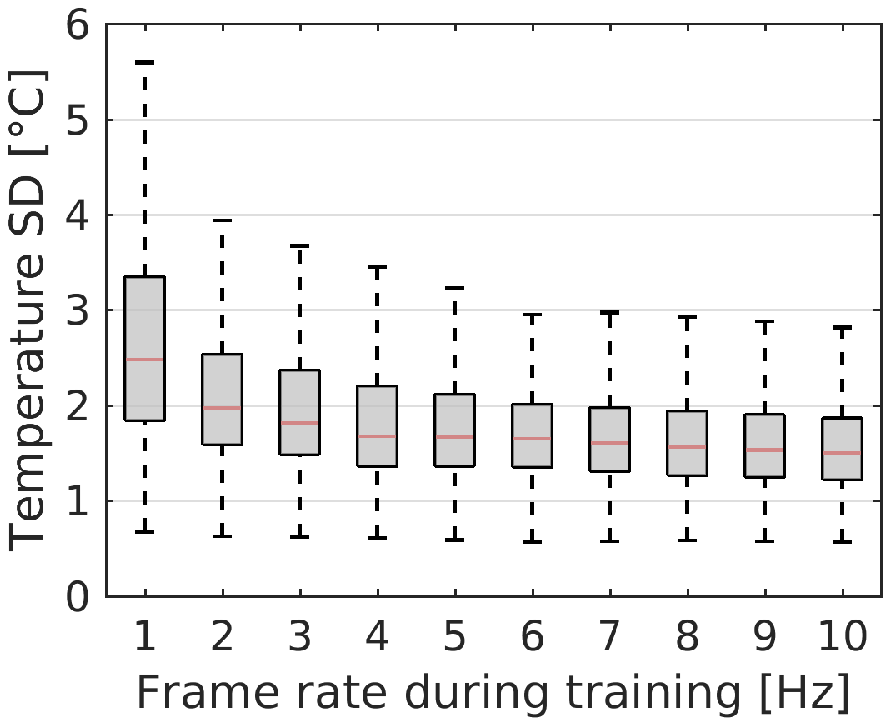}}
\centerline{(a)}\medskip
\end{minipage}
\begin{minipage}[b]{0.49\linewidth}
\centerline{\large{Temperature accuracy}}\medskip
\vspace{-0.2cm}
\centerline{\includegraphics[trim={0cm 0cm 0cm 0cm},clip,height=5.5cm]{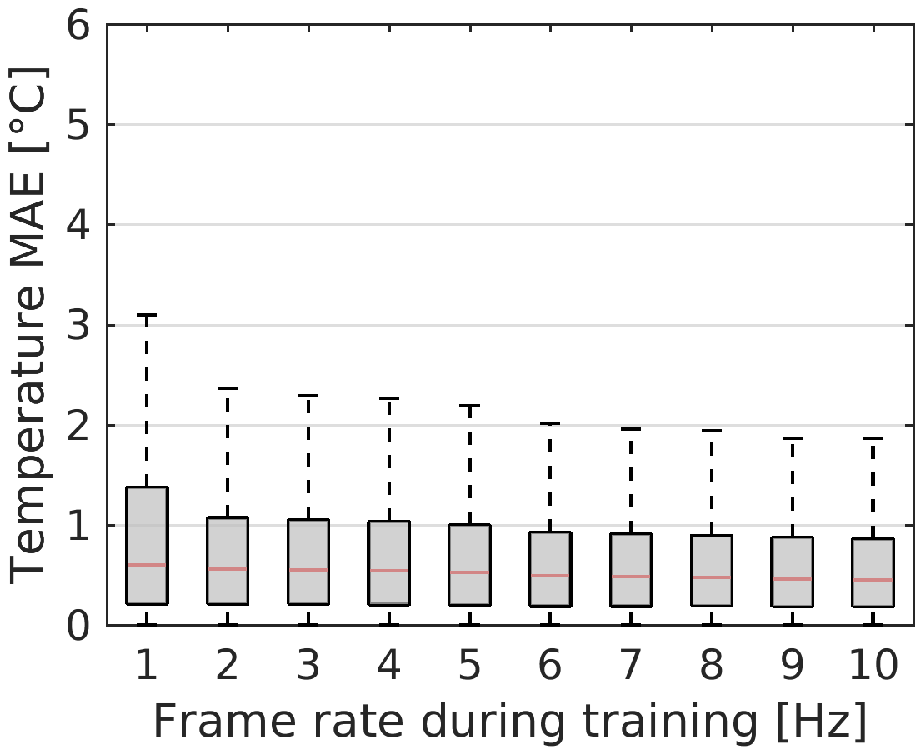}}
\centerline{(b)}\medskip
\end{minipage}
\caption{Analysis of the impact of the amount of learning images on the temperature stability. The temperature precision (a) and accuracy (b) were evaluated using various imaging frame rates for training. Box-and-whisker plots of the temporal temperature standard deviation (a) and mean absolute error (b) obtained over the 12 volunteers are reported for imaging frame rates ranking between 1 and 10 Hz. Note that the number $N$ of dynamic images used to train the CNN-model was equal to 20 (resp. 40, 60, ..., 200) when an imaging frame rate of 1 Hz (resp. 2, 3, ... 10 Hz) was employed.}
\label{fig:framerate}
\end{figure}

Figure \ref{fig:framerate} analyzes the impact of the amount of learning images: compared to the original 10 Hz imaging frame rate, a 2 Hz frame rate deteriorated moderately thermometric precision and accuracy (by less than $30 \%$).

\subsection{\textit{In vivo} heating study in a porcine liver}

\begin{figure}[h!]

\begin{minipage}[b]{0.25\linewidth}
\centering
\vspace{-1.5cm}
 \centerline{\large{Thermal-map}}\medskip
 \centerline{\large{$t=100$ s}}\medskip
\vspace{1.5cm}
\end{minipage}
\begin{minipage}[b]{0.2\linewidth}
\centerline{MB}\medskip
\centerline{\includegraphics[trim={0.3cm 1cm 1cm 1cm},clip,width=2.5cm]{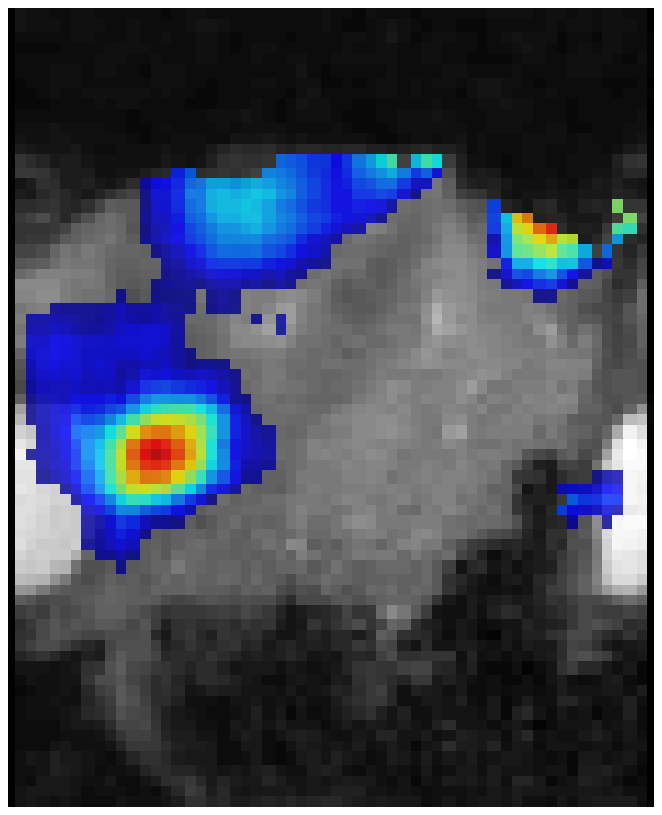}}
\centerline{(a)}\medskip
\end{minipage}
\begin{minipage}[b]{0.2\linewidth}
\centerline{LM}\medskip
\centerline{\includegraphics[trim={0.3cm 1cm 1cm 1cm},clip,width=2.5cm]{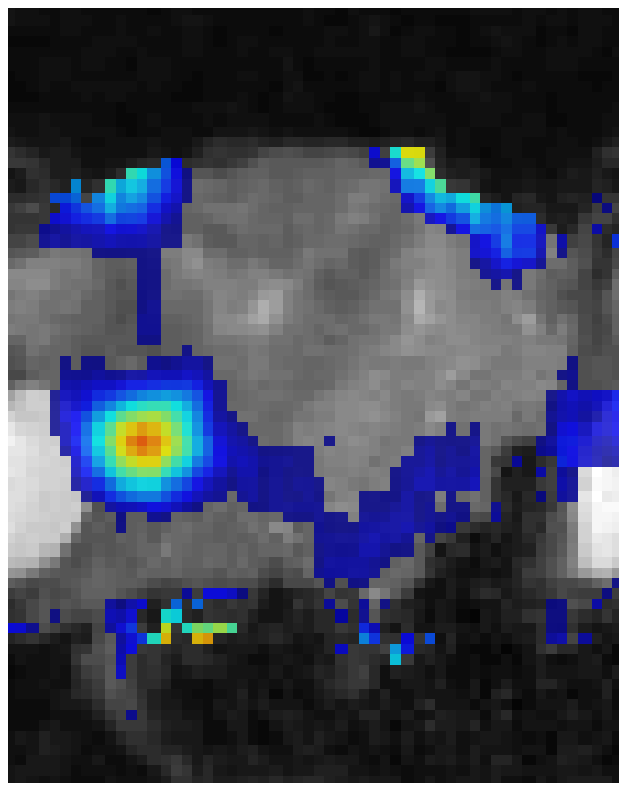}}
\centerline{(b)}\medskip
\end{minipage}
\begin{minipage}[b]{0.2\linewidth}
\centerline{CNN}\medskip
\centerline{\includegraphics[trim={0.3cm 1cm 1cm 1cm},clip,width=2.5cm]{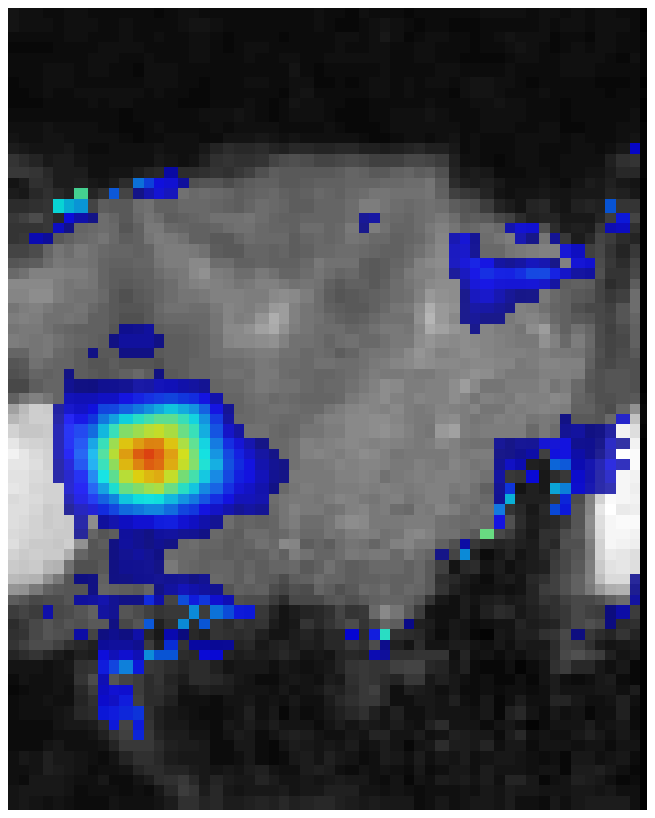}}
\centerline{(c)}\medskip
\end{minipage}
\begin{minipage}[b]{0.08\linewidth}
\centerline{\includegraphics[trim={0cm 0cm 0cm 0cm},clip,height=2.5cm]{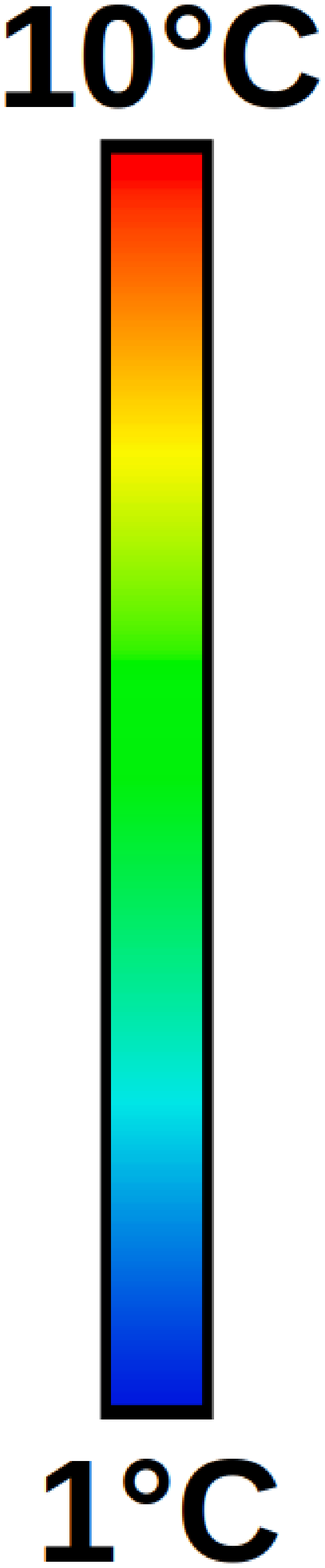}}
\centerline{}\medskip
\end{minipage}

\begin{minipage}[b]{0.25\linewidth}
\centering
\vspace{-2cm}
 \centerline{\large{SD-map}}\medskip
\vspace{2cm}
\end{minipage}
\begin{minipage}[b]{0.2\linewidth}
\centerline{MB}\medskip
\centerline{\includegraphics[trim={0.2cm 0.6cm 0.5cm 0.6cm},clip,width=2.5cm]{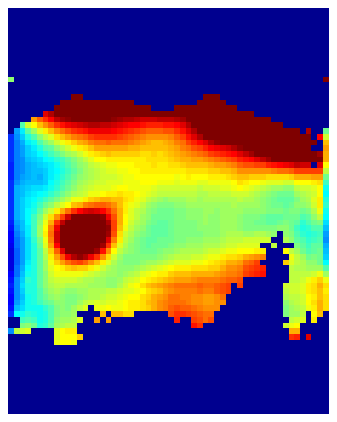}}
\centerline{(d)}\medskip
\end{minipage}
\begin{minipage}[b]{0.2\linewidth}
\centerline{LM}\medskip
\centerline{\includegraphics[trim={0.2cm 0.5cm 0.5cm 0.5cm},clip,width=2.5cm]{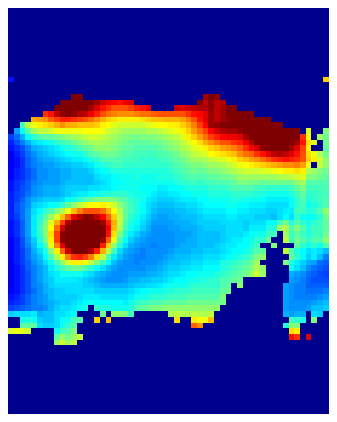}}
\centerline{(e)}\medskip
\end{minipage}
\begin{minipage}[b]{0.2\linewidth}
\centerline{CNN}\medskip
\centerline{\includegraphics[trim={0.2cm 0.5cm 0.5cm 0.5cm},clip,width=2.5cm]{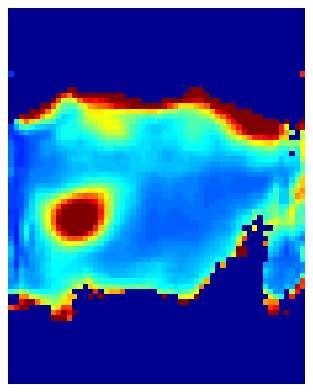}}
\centerline{(f)}\medskip
\end{minipage}
\begin{minipage}[b]{0.08\linewidth}
\centerline{\includegraphics[trim={0cm 0cm 0cm 0cm},clip,height=3cm]{./Figures/error_bar.eps}}
\centerline{}\medskip
\end{minipage}

\begin{minipage}[b]{0.25\linewidth}
\centering
\vspace{-2cm}
 \centerline{\large{MAE-map}}\medskip
\vspace{2cm}
\end{minipage}
\begin{minipage}[b]{0.2\linewidth}
\centerline{MB}\medskip
\centerline{\includegraphics[trim={0.2cm 0.5cm 0.5cm 0.5cm},clip,width=2.5cm]{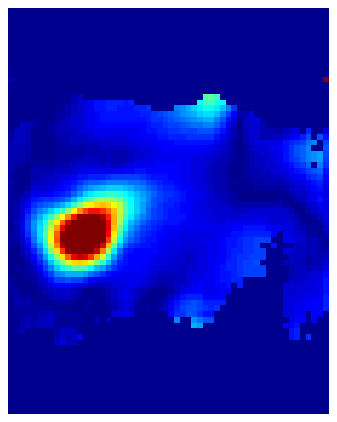}}
\centerline{(h)}\medskip
\end{minipage}
\begin{minipage}[b]{0.2\linewidth}
\centerline{LM}\medskip
\centerline{\includegraphics[trim={0.2cm 0.5cm 0.5cm 0.5cm},clip,width=2.5cm]{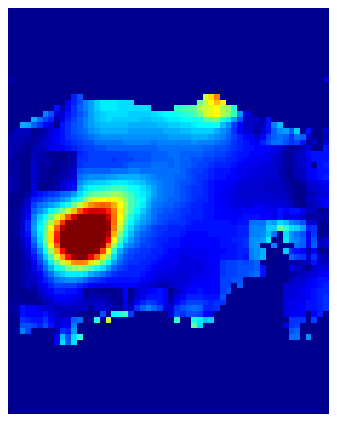}}
\centerline{(i)}\medskip
\end{minipage}
\begin{minipage}[b]{0.2\linewidth}
\centerline{CNN}\medskip
\centerline{\includegraphics[trim={0.2cm 0.5cm 0.5cm 0.5cm},clip,width=2.5cm]{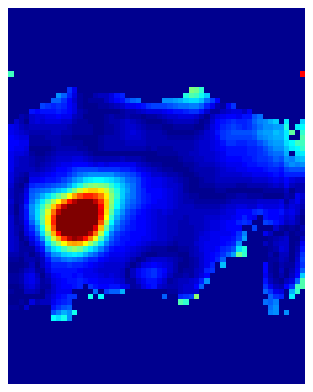}}
\centerline{(j)}\medskip
\end{minipage}
\begin{minipage}[b]{0.08\linewidth}
\centerline{\includegraphics[trim={0cm 0cm 0cm 0cm},clip,height=3cm]{./Figures/error_bar.eps}}
\centerline{}\medskip
\end{minipage}

\begin{minipage}[b]{0.32\linewidth}
\centerline{MB}\medskip
\centerline{\includegraphics[trim={0cm 0cm 0cm 0cm},clip,height=3.75cm]{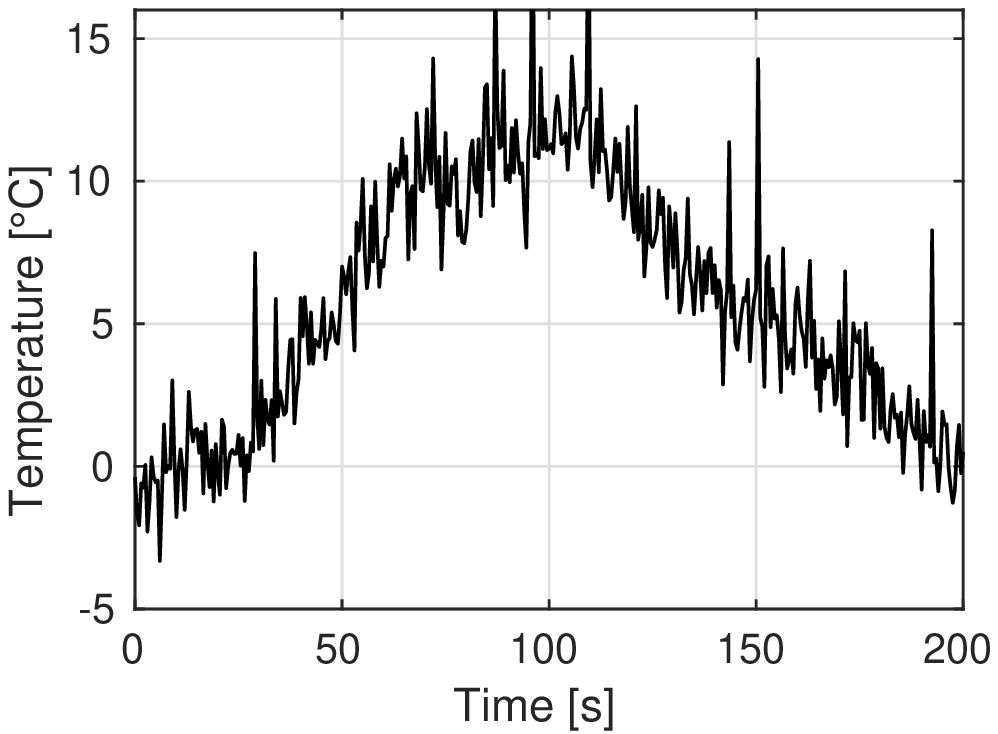}}
\centerline{(k)}\medskip
\end{minipage}
\begin{minipage}[b]{0.32\linewidth}
\centerline{LM}\medskip
\centerline{\includegraphics[trim={0cm 0cm 0cm 0cm},clip,height=3.75cm]{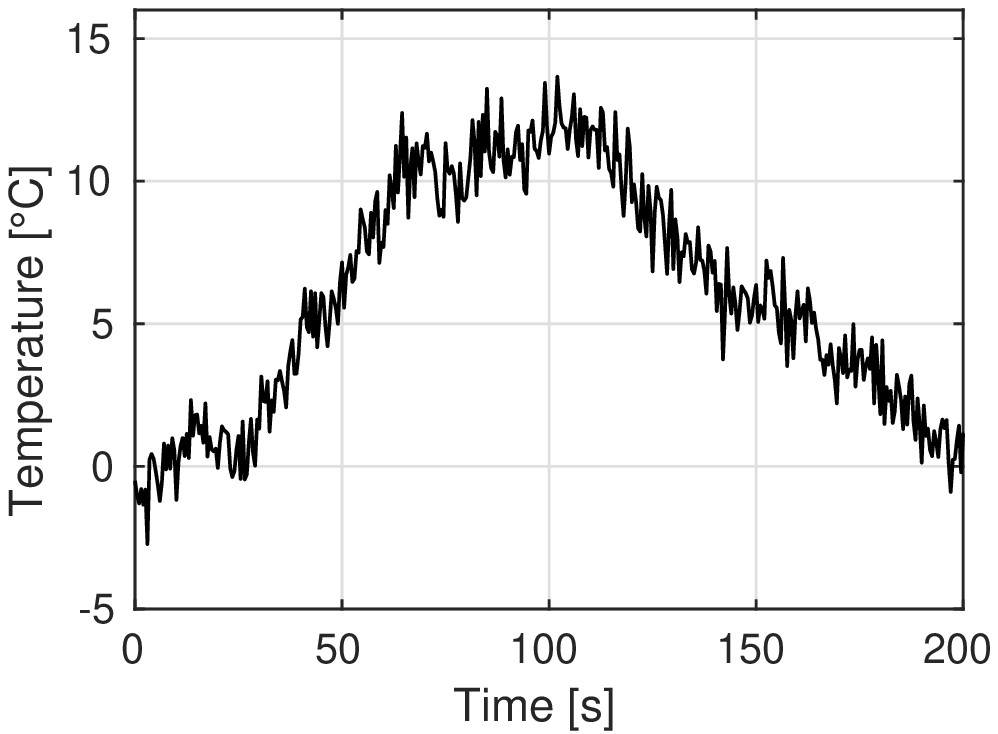}}
\centerline{(l)}\medskip
\end{minipage}
\begin{minipage}[b]{0.32\linewidth}
\centerline{CNN}\medskip
\centerline{\includegraphics[trim={0cm 0cm 0cm 0cm},clip,height=3.75cm]{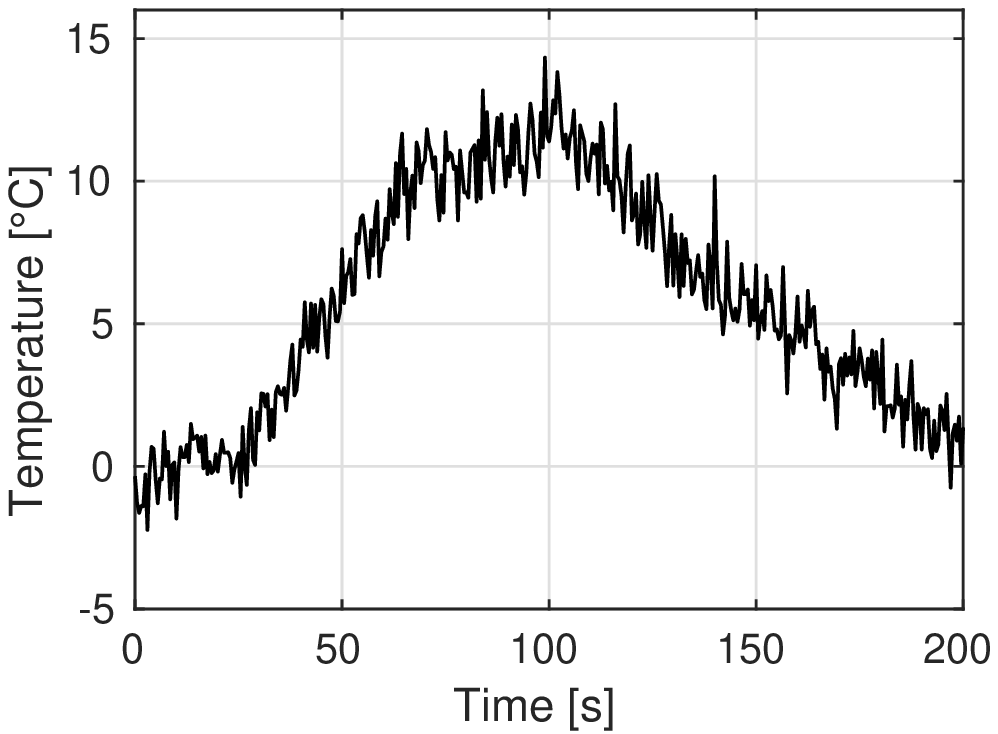}}
\centerline{(m)}\medskip
\end{minipage}

\caption{MR thermometry results obtained in a porcine liver during HIFU heating: (first row) Temperature maps obtained after 80 s of heating ($t=100 s$) overlaid on the anatomic image, (second row) temporal standard deviation map, (third row) mean absolute error map, and (bottom row) temporal evolution of the temperature in a single voxel located at the focal point position. Results are reported using LUT (left), LM (middle) and CNN (right).}
\label{fig:thermo_pig}
\end{figure}

Figure \ref{fig:thermo_pig} shows MR thermometry results obtained on a porcine liver during HIFU heating. Thermal maps are reported after 80 s of sonication using LUT (\ref{fig:thermo_pig}a), LM (\ref{fig:thermo_pig}b) and CNN (\ref{fig:thermo_pig}c). In absence of any correction strategy, apparent temperature fluctuations of up to 13$^{\circ}$C (peak-to-peak) were observed in the target area. The heated region using the LUT appears slightly elongated as compared to LM and CNN. Residual thermometry artifacts are observable in the upper part of the liver with LUT (see \ref{fig:thermo_pig}a). These apparent temperature fluctuations are however greatly reduced using LM (\ref{fig:thermo_pig}b), and even more using CNN (\ref{fig:thermo_pig}c). This visual observation is confirmed in associated SD-maps: in most of the voxels located in the upper part of the liver, a temperature standard deviation higher than 3$^{\circ}$C using LUT (\ref{fig:thermo_pig}d) decreased until 2$^{\circ}$C using LM (\ref{fig:thermo_pig}e), and reached 1$^{\circ}$C using CNN (\ref{fig:thermo_pig}f). This precision gain with CNN was achieved without creating any additional offset, as shown by the MAE-maps (see \ref{fig:thermo_pig}h, \ref{fig:thermo_pig}i and \ref{fig:thermo_pig}j). The evolution of the temperature is shown in a single voxel located at the focal point position using LUT (\ref{fig:thermo_pig}k), LM (\ref{fig:thermo_pig}l) and CNN (\ref{fig:thermo_pig}m). Higher residual temporal temperature fluctuations are observable using the LUT correction as compared to the other two correction approaches. LM and CNN approaches lead to a comparable observation of the temperature evolution: a temperature increase of 12$^\circ$C was reached after 80 s of HIFU sonication.

\subsection{Benchmark}
\label{ssec:benchmark}

\begin{figure}[h!]
\begin{minipage}[b]{0.49\linewidth}
\centerline{\large{No fine-tuning}}\medskip
\centerline{\includegraphics[trim={0cm 0cm 0cm 0cm},clip,height=5.5cm]{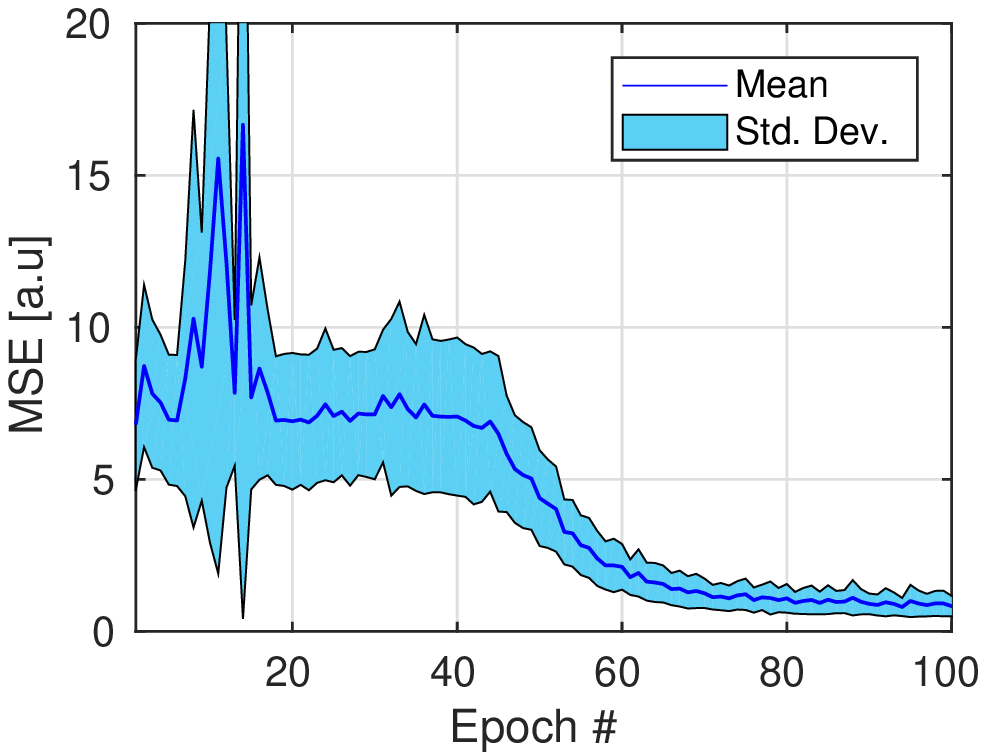}}
\centerline{(a)}\medskip
\end{minipage}
\begin{minipage}[b]{0.49\linewidth}
\centerline{\large{With fine-tuning}}\medskip
\centerline{\includegraphics[trim={0cm 0cm 0cm 0cm},clip,height=5.5cm]{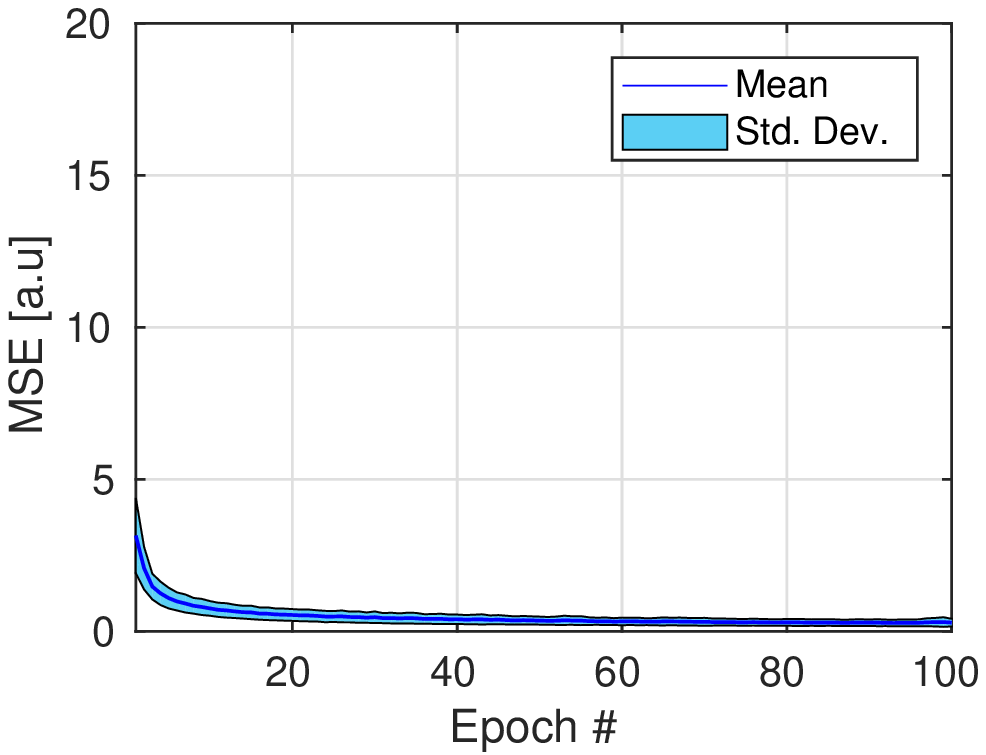}}
\centerline{(b)}\medskip
\end{minipage}
\caption{Loss metric as a function of the number of epochs obtained without (a) and with (b) the use of fine-tuning. For each number of epochs, the mean and standard deviation of 13 MSE values (\emph{i.e.}, 12 values for the volunteer study + 1 value for the heating study) are reported.}
\label{fig:FT}
\end{figure}

During the learning stage, around $1.5$ s were required in average for the accomplishement of one epoch. Figure \ref{fig:FT} shows the loss metric as a function of the number of epochs without and with the proposed fine-tuning strategy. It can be observed that the use of fine-tuning stabilized and accelerated the convergence of the optimization process, and this for all data sets involved in this study.

During the hyperthermia session, 150 ms were required to generate one single motion compensated temperature map. In such a case, the GPU usage was however highly under-exploited: interestingly, the calculation of a pack of $\delta=$16 maps could be also accomplished within 150 ms.

\section{Discussion}

The proposed method is designed to remove motion-related susceptibility effects induced by breathing in real-time abdominal MR-thermometry. To this end, the existing multi-baseline strategy is extended using a deep neural network: a CNN learns the apparent temperature perturbation during a preparative learning stage performed before hyperthermia.

A major difficulty when dealing with deep learning is the inherent computational cost. One goal of this study was to investigate if (i) the amount of learning images and (ii) the CNN training time can be optimized to the point that clinical thermotherapy interventions are feasible. For this, all calculations at the different stages of the proposed workflow (\emph{i.e.}, learning stage, CNN model optimization, determination of an offset map and interventional session) were designed to be compatible with the constraints of clinical thermotherapy procedures (see section \ref{ssec:benchmark}). With respect to the learning stage, 20 epochs using fine-tuning provide a loss similar to the one achieved by 100 epochs (see figure \ref{fig:FT}). The calculation of the CNN-model (red block in figure \ref{fig:CNN_scheme}) could thus be accomplished within less than 30 s (resp. less than 6 s) using our test platform with a training frame rate of 10 Hz (resp. 2 Hz). The compensation of the time-persistent offset (as described in section \ref{sssec:bias}), which relied on a CNN-correction for each of the $N$ training images, could be accomplished within less than a second in all presented experiments. Regarding the interventional session, it is imperative that all calculations have to be done within the interval of subsequent image aquisitions in order to prevent back-log. The use of a sliding temporal window (as introduced in section \ref{sssec:temp_win}) of size $\delta=2$ dynamic was mandatory in the volunteer study to cope with a 10 Hz imaging frame rate.

A second major challenge is the presence of noise and wraps in MR-phase images, which hampers the CNN optimization process (red block in figure \ref{fig:CNN_scheme}). This drawback was first partially addressed by the use of temperature maps as inputs for the CNN instead of the phase images. However, a voxelwise time-persistent offset remained, induced by the the presence of noise in the baseline phase image $\varphi (\vec{r},t_0)$ in Eq. (\ref{eq:PRF}). This issue was addressed using an offset correction (as described in section \ref{sssec:bias}), and no additional penalty in the accuracy was observable using CNN as compared to LUT and LM. 

Using LM, a linear phase model is derived from the resolution of an overdetermined system of $N$ reference images. In comparison with LUT, noise may be reduced on the resynthesized baseline phase image in Eq. (\ref{eq:PRF}). Ideally, a ``noise-free'' resynthesized baseline phase images is produced, and the noise contribution on temperature uncertainty is reduced by a factor $\sqrt{2}$. Moreover, while LUT intrinsically cannot correct for motion amplitudes higher than the ones observed in the learning stage, LM can still provide an extrapolation of the reference phase. As a consequence, regarding the precision of MR-thermometry, LM outperformed LUT in all presented experiments. Our findings show that the multi-baseline strategy can be further improved using CNN. It must be underlined that an inherent drawback with CNN lies in the high complexity of the fitted model, which makes it difficult to interpret. However, CNN was able to cope with complex motion-related thermometry artifacts (as encountered in the upper part of the liver), for which an explicit modeling is very challenging on-line.

The CNN method provided accuracte temperature measurements for imaging frame rate above 2 Hz. The proposed CNN method was also demonstrated to be compatible with fast MR acquisition schemes of up to 10 Hz. Real-time MR-thermometry may thus be advantageously combined with any suitable real-time temporal filtering to further improve the measurement precision, as described in \cite{pipeline} or \cite{NLM_thermo}.
 
It should be noted that LUT and LM rely on the on-line determination of a motion surrogate, which can be provided by various types of sensors such as breathing belt or MR-/ultrasound-based surrogates. Using CNN, no motion surrogate is required since motion patterns are implicitely extracted from the actual magnitude image. CNN thereby provides an independent thermal information with that provided by LUT and LM (among others), which opens great perspectives for the use of the method as a ``Watchdog'' for on-line quality control (QC).

Several other machine learning models have been considered (such as Logistic Regression, Naïve Bayes, Random Forest and Support Machine Vector) to learn motion related errors in abdominal MR-thermometry in the current study. However, using such algorithms, the computation cost was consistently much higher than LM, which relies on a very simple linear model. Moreover, the above-mentioned machine learning models showed difficulties to interpolate/extrapolate positions not observed during training. The LM model, as implemented in the current study, appeared more appropriate for this task, since the pixel wise temperature variation with respiratory motion in the abdomen can be efficiently approximated in first order with a linear term, as shown in \cite{LM_MICCAI_2007}. As a consequence, we decided to limit the scope of this paper to a comparison of the two already published multi-baseline methods: LUT and LM.

The main limitation of the proposed method --- as it is common with multi-baseline strategies --- is its inability to compensate for thermometry artifacts related to motion / deformation(s), which has not been observed during the training period. In practice, this can be encountered due to physiological drift or spontaneous motion. If during hyperthermia bulk shofts or major drifts from the calibration position are observed, a recalibration of the correction data is then required.

\section{Conclusion}

PRF-based MR-thermometry is complicated in abdominal organs by displacement of the target and surrounding tissues, which hampers direct voxel-by-voxel comparisons. Strong temperature artifacts are introduced by motion-induced additional phase variations via an inhomogeneous and time-variant magnetic field. The proposed approach extends the existing multi-baseline strategy using CNN to address such artifacts in abdominal organs due to breathing. A workflow is proposed to solve inherent issues with CNN associated to computational burden for training. The proposed method outperformed two existing multi-baseline strategies in terms of temperature precision, especially in regions prone to strong susceptibility artifacts as encountered in the upper part of the liver. This was achieved without noteworthy additional penalty in the temperature accuracy. Moreover, we have demonstrated that, even under clinical conditions, the method was found robust and artifact-free in all examined cases and well able to follow the temperature evolution of an \textit{in vivo} HIFU ablation.

\section*{Acknowledgment}

Experiments presented in this paper were carried out using the PlaFRIM experimental testbed, supported by Inria, CNRS (LABRI and IMB), Universit\'e de Bordeaux, Bordeaux INP and Conseil R\'egional d'Aquitaine (see https://www.plafrim.fr/). Computer time for this study was provided by the computing facilities MCIA (M\'esocentre de Calcul Intensif Aquitain) of the Universit\'e de Bordeaux and of the Universit\'e de Pau et des Pays de l'Adour.

\section*{References}

\bibliographystyle{dcu}
\bibliography{2020_Deep_Thermometry_clean_copy}

\begin{table}
\footnotesize
\begin{tabular*}{\textwidth}{@{}c*{6}{ccc}}
\cline{1-3}\\
\multicolumn{3}{c}{\textbf{Implemented CNN model}}\\
\cline{1-3}\\
\textbf{Layer type (Id [$\#$])} & \textbf{Output shape} & \textbf{Number of parameters [$\#$]} \\
\cline{1-3}\\
InputLayer (1) &           (128, 128, 1) & 0 \\
BatchNormalization (1) & (128, 128, 1) & 4 \\
Conv2D (1) &              (128, 128, 24) & 240 \\
BatchNormalization (2) & (128, 128, 24) & 96 \\
Conv2D (2) &              (128, 128, 48) & 10416 \\
MaxPooling2D (1) & (64, 64, 48)  & 0 \\
Dropout (1) &            (64, 64, 48) &  0 \\
BatchNormalization (3) & (64, 64, 48) &  192 \\
Conv2D (3) &              (64, 64, 48) &  20784 \\
BatchNormalization (4) & (64, 64, 48) &  192 \\
Conv2D (4) &              (64, 64, 96) &  41568 \\
MaxPooling2D (2) & (32, 32, 96) &  0 \\
Dropout (2) &            (32, 32, 96) &  0 \\
BatchNormalization (5) & (32, 32, 96) &  384 \\
Conv2D (5) &              (32, 32, 96) &  83040 \\
BatchNormalization (6) & (32, 32, 96) &  384 \\
Conv2D (6) &              (32, 32, 192) & 166080 \\
MaxPooling2D (3) & (16, 16, 192) & 0 \\
Dropout (3) &            (16, 16, 192) & 0 \\
BatchNormalization (7) & (16, 16, 192) & 768 \\
Conv2D (7) &              (16, 16, 192) & 331968 \\
BatchNormalization (8) & (16, 16, 192) & 768 \\
Conv2D (8) &              (16, 16, 384) & 663936 \\
UpSampling2D (5) &              (32, 32, 384) & 0 \\
Concatenate (1) &    (32, 32, 576) & 0 \\
BatchNormalization (9) & (32, 32, 576) & 2304 \\
Conv2D (9) &              (32, 32, 192) & 995520 \\
BatchNormalization (10) & (32, 32, 192) & 768 \\
Conv2D (10) &             (32, 32, 96) &  165984 \\
UpSampling2D (6) &              (64, 64, 96) &  0 \\
Concatenate (2) &    (64, 64, 192) & 0 \\
BatchNormalization (11) & (64, 64, 192) & 768 \\
Conv2D (11) &             (64, 64, 96) &  165984 \\
BatchNormalization (12) & (64, 64, 96) &  384 \\
Conv2D (12) &             (64, 64, 48) &  41520 \\
UpSampling2D (7) &              (128, 128, 48) & 0 \\
Concatenate (3) &    (128, 128, 96) & 0 \\
BatchNormalization (13) & (128, 128, 96) & 384 \\
Conv2D (13) &             (128, 128, 48) & 41520 \\
BatchNormalization (14) & (128, 128, 48) & 192 \\
Conv2D (14) &             (128, 128, 24) & 10392 \\
Conv2D (15) &             (128, 128, 1) & 217 \\
\cline{1-3}\\
\multicolumn{2}{r}{\textbf{Total number of parameters}} & \textbf{2746757} \\
\cline{1-3}\\
\end{tabular*}
\caption{Output shape and number of parameters involved in each layer of the implemented CNN-model for a $128 \times 128$ image.}
\label{table:model}
\end{table}

\end{document}